\documentclass[twocolumn]{aastex63}

\newcommand{\kms}{km s$^{-1}$}

\usepackage{color}
\usepackage{multirow}
\usepackage{amsmath}
\usepackage{textcomp}
\usepackage{booktabs}
\usepackage{natbib}
\newcommand{\uat}[2]{\href{http://astrothesaurus.org/uat/#2}{#1 (#2)}}
\accepted{February 24, 2022}
\submitjournal{ApJ}

\shorttitle{NGC\,4437 Group \& Magnitude Gap}
\shortauthors{Kim et al.}

\begin{document}

\title{ 
A Rich Satellite Population of the NGC\,4437 Group and \\
Implications of a Magnitude Gap for Galaxy Group Assembly History %\\
%{\bf (Revised, \today)}
}

\correspondingauthor{Myung Gyoon Lee}
\email{yoojkim@astro.snu.ac.kr, mglee@astro.snu.ac.kr}

\author[0000-0003-1392-0845]{Yoo Jung Kim} 
\author[0000-0003-3734-1995]{Jisu Kang} 
\author[0000-0003-2713-6744]{Myung Gyoon Lee} 
\affiliation{Astronomy Program, Department of Physics and Astronomy, Seoul National University, 1 Gwanak-ro, Gwanak-gu, Seoul 08826, Republic of Korea}

%\email{mglee@astro.snu.ac.kr}
\author[0000-0002-2502-0070]{In Sung Jang} 
\affiliation{Department of Astronomy \& Astrophysics, Univ. Chicago, 5640 S. Ellis Avenue, Chicago, IL 60637, USA}

% ===================================
%     Abstract
% ===================================
\begin{abstract}
{
Both observations and cosmological simulations have recently shown that there is a large scatter in the number of satellites of Milky Way (MW)-like galaxies. 
In this study, we investigate the relation between the satellite number and galaxy group assembly history, using %\textcolor{blue}{the} 
the $r-$band 
magnitude gap ($\Delta m_{12}$) between the first and the second brightest galaxy as an indicator. 
From 20 deg$^2$ of Hyper Suprime-Cam Subaru Strategic Program %(HSC-SSP)
Wide layer, we identify 17 dwarf satellite candidates around NGC\,4437, a spiral galaxy with about one-fourth of the MW stellar mass. %, paired with the 2.5 mag fainter dwarf spiral galaxy NGC\,4592. 
We estimate their distances %of the candidates 
using the surface brightness fluctuation (SBF) method. Then we confirm five candidates as members of the NGC\,4437 group, resulting in a total of seven %\textcolor{blue}{group}
group members. 
%While \textcolor{blue}{the satellite population of the NGC\,4437 group} is relatively rich among observed and simulated galaxy groups with similar host stellar masses, it is in the expected range when compared with groups with similar $\Delta m_{12}$. 
Combining the NGC\,4437 group (with $\Delta m_{12} = 2.5$ mag) with other groups in the literature,
we find a stratification of the satellite number %of satellites 
by $\Delta m_{12}$ for a given host stellar mass. 
The satellite number %of satellites 
for given host stellar mass decreases as $\Delta m_{12}$ increases.
The same trend is found in simulated galaxy groups in IllustrisTNG50 %cosmological
simulations. 
%From these simulated galaxy groups, 
We also find that the host galaxies in groups with a smaller $\Delta m_{12}$ (like NGC\,4437) have assembled their halo mass more recently than those in larger gap groups, and that their stellar-to-halo mass ratios (SHMRs)  increase as $\Delta m_{12}$ increases.
These results show that the large scatter in the satellite number %of satellites 
is consistent with a large range of $\Delta m_{12}$, indicating diverse group assembly histories.
%Thus $\Delta m_{12}$ is an efficient indicator to trace galaxy group assembly history.
}

\end{abstract}
%========================================
\keywords{\uat{Galaxy Groups}{597}; \uat{Dwarf galaxies}{416};
\uat{Galaxy dark matter halos}{1880};\uat{Distance indicators}{394}; \uat{Galaxy distances}{590}; \uat{Luminosity Function}{942}}

% ===================================
%     1. Introduction
% ===================================

\section{Introduction}\label{sec_introduction}

%Searching for satellite galaxies around central galaxies has been an important goal in observational cosmology. 
In the $\Lambda$CDM paradigm, galaxies form in the center of dark matter halos and evolve through hierarchical merging and accretion. 
%\textbf{
Smaller halos that are accreted into larger halos and become gravitationally bound are called subhalos, which observationally correspond to the satellite galaxies. %}
%Smaller halos %\textcolor{blue}{}
%that are accreted into larger halos and become gravitationally bound %, and they 
%are called satellite galaxies (subhalos). 
%Therefore, it 
It is inferred that the more massive the dark matter halos are, the more abundant the satellite systems are.

This structure formation model has been tested by comparing observations with simulations.
The well-known “Missing Satellites” problem \citep{kly99, mor99} is an example. 
With dark matter-only cosmological simulations, Milky Way (MW)-like halos in the $\Lambda$CDM scenario are predicted to have far more satellites than %\textcolor{blue}{
observed around the MW and %} % and} %observed satellites around the MW, which 
it is considered as one of the “small-scale” challenges to $\Lambda$CDM \citep[see][and references therein]{bul17}. 
This motivated numerous studies of individual satellite systems, % of MW-like galaxies, 
to find out whether the MW satellite population is representative among them,
from the Local Volume (LV; D $<$ 11Mpc) host galaxies (e.g. M31 \citep[see][for compiled data]{mar16, mcc12, mcc18}, %\textcolor{red}{
M81 \citep{chi09, chi13},
NGC\,5128 (Centaurus A) \citep{crn14, cnr19, mul17, mul19}, 
M94 \citep{sme18}, M101 \citep{dan17, ben17, ben19, car19a}, 
NGC\,4258 (M106) \citep{spe14, kim11},  %{\color{red} (Kim+11?, MNRAS, 412, 1881)}
NGC\,628 \citep{dav21}) to those in farther distances (NGC\,3175 \citep{kon18}, NGC\,2950 and NGC\,3245 \citep{tan18}).

In more recent years, statistical studies of satellites around MW-like galaxies have been carried out \citep{car20a, car20b, car21, wan21, geh17, hab20, mao21, rob21}. 
Among them \citet{car21} presented a study of a large sample of dwarf satellite systems around 12 LV host galaxies within the projected distance to their host galaxy ($R_h$) of 150 kpc, to which surface brightness fluctuation (SBF) or the tip of the red giant branch (TRGB) distances are measured. They found that the MW satellite luminosity function is typical among other MW-like LV hosts. 

At farther distances beyond the LV, while it is more difficult to identify and confirm membership of faint galaxies, clear advantages exist: there are a larger number of MW-like galaxies and a smaller field of view is sufficient to survey the entire virial volume. 
Among others, the SAGA survey \citep{geh17, mao21} presented a study of spectroscopically confirmed satellites with a  homogeneous spatial and photometric completeness. The SAGA Stage II \citep{mao21} presented 127 satellites within $R_h = 300$ kpc around 36 MW-like hosts at distance 25 Mpc $< D <$ 40.75 Mpc. 
They suggested that the number of satellites ($M_r < -12.3$ mag) of the MW-like hosts is remarkably varied, from zero to nine.
In their sample, the MW has a typical number of satellites.

Meanwhile, several theoretical studies found that baryonic cosmological simulations produce fewer observable satellites around MW-like hosts
compared with previous simulation studies, which is consistent with observations  \citep{wet16, gar19, fat16, saw16, fon21}. 
Also, the diversity of satellite number was shown by \citet{eng21b} %that 
who studied satellite systems of 198 MW-like hosts found in IllustrisTNG50 \citep{nel19}. 

Thus, there is growing consensus between observations and simulations 
that the number of satellites of the MW is typical among other observed and simulated systems.
%\textcolor{blue}{
Besides, a large \emph{scatter} in the number of satellites is recognized from both approaches. % better word than besides?

One of the causes of the large scatter might be \emph{diverse galaxy assembly histories}. In the hierarchical structure formation scenario, it is expected that an early-formed system would have already cannibalized its massive satellite galaxies, if existed. Therefore, the early-formed system would likely have a small number of satellite galaxies with the central galaxy dominating the brightness of the group.
Indeed, \citet{mao21} noted that MW-mass galaxy groups with massive satellites in the SAGA survey have a larger number of satellites. In addition, \citet{sme21} found a tight relationship between the mass of a galaxy's most dominant merger and its number of satellites, using stellar halo and satellite properties of eight nearby MW-like galaxies. As a metric for the galaxy's most dominant merger, they used either the total accreted stellar mass or the mass of the most massive satellite.

In this view, a %\textcolor{blue}{
magnitude difference, or a %}
\emph{magnitude gap} ($\Delta m_{12}$) between the brightest and the second brightest galaxy %\textcolor{blue}{
gives useful information about %} %can inform 
galaxy assembly history. A galaxy group with a large magnitude gap is likely to have assembled its mass at an early epoch. Indeed, fossil groups, which are massive groups observationally selected on the basis of $r-$band magnitude gap, $\Delta m_{12}>2$ mag and X-ray luminosity $L_X > 10^{42}h^{-2}_{50} {\rm erg s^{-1}}$ \citep{jon03}, are considered to have assembled their mass earlier than non-fossil groups \citep{don05} or have lacked recent infall of new massive satellites \citep{kun17}  
(See \citet{agu21} for a recent review on fossil groups).
Note that $\Delta m_{12}$ is largely varied among nearby galaxy groups. 
While the M81 group has a massive satellite M82 and thus 
has a small gap, $\Delta m_{12}=1.4$ mag \citep[calculated from][]{dev91}, the M94 group has only faint satellites in its virial volume and thus %have 
has a very large gap, $\Delta m_{12}=9.9$ mag \citep{sme18}.

However, only a few studies have examined the observational evidence of the anticorrelation between the magnitude gap and the number of satellites 
\citep{hea13, wan21}.
\citet{hea13} found that SDSS groups with small magnitude gaps ($\Delta m_{12}<0.2$ mag)  
have a richer satellite system than large-gap groups ($\Delta m_{12}>1.5$ mag) at fixed %\textbf{
velocity dispersion among group members as a mass proxy
%}
, at about MW-mass range. They used a volume-limited galaxy group catalog complete down to absolute $r-$band magnitude $M_r = -19$ mag
identified in Data Release 7 of the SDSS. 
Therefore, their sample does not contain large-gap groups with the second brightest galaxy fainter than $M_r = -19$ mag. This criteria would exclude most of the nearby galaxy groups with low mass. 
In addition, \citet{wan21} suggested that MW-mass galaxy groups with magnitude gaps smaller than one mag have richer satellite systems than those with larger magnitude gaps.
However, most nearby galaxies have magnitude gaps larger than one mag.
Therefore, to understand any relation between the magnitude gap %with 
and a large scatter in satellite number among nearby MW-like galaxies, more observations of nearby satellite systems with various magnitude gaps as well as comparisons with simulations are required.

Here we study satellite populations of the NGC\,4437 group, which has a relatively small magnitude gap ($\Delta m_{12} = 2.5$ mag) among nearby galaxy groups but larger than the previously studied regime. Then we compare the NGC\,4437 group with those of other nearby galaxy groups as well as with mock galaxy groups from cosmological simulations.

NGC\,4437 (=NGC\,4517) is a LV spiral galaxy ($M_r = -20.71$ mag, $D=9.28\pm0.39$ Mpc) with about one fourth the stellar mass of the MW, forming a pair with NGC\,4592 which is a Large Magellanic Cloud (LMC)-mass dwarf spiral galaxy ($M_r = -18.23$ mag, $D=9.07\pm0.27$ Mpc)  \citep{kim20}. See Table \ref{tab_spirals} for the basic information of the two spiral galaxies. 
Although various surveys have detected several faint galaxies around NGC\,4437 and NGC\,4592, their relationship with these spiral galaxies %were 
was not defined except CGCG 014-054, a late-type dwarf galaxy at a Tully-Fisher distance $D = 9.59 \pm 1.77$ Mpc that was grouped together with the two galaxies by \citet{kar13a}.

In this study, we surveyed a wide area around NGC\,4437 in the WIDE layer of Hyper Suprime-Cam Subaru Strategic Program \citep[HSC-SSP,][]{aih18} 
and found 17 dwarf satellite candidates.
To distinguish them from background galaxies in the Virgo cluster ($\sim 16.5$ Mpc \citep{mei07}), we %confirmed
checked their membership by measuring their SBF distances.
In doing so, we present a spatially complete list of confirmed satellite galaxies around NGC\,4437, complete down to $M_r \sim -11$ mag.

This paper is structured as follows.
In Section \ref{sec_data}, we describe the HSC data and our detection of dwarf satellite candidates.
In \S3, %\ref{sec_sbf}, 
we present SBF distances to the dwarf satellite candidates and discuss their membership %of 
%\textcolor{blue}{to/in} 
in the NGC\,4437 group.
In \S\ref{sec_beyond}, we describe the spatial distribution of the NGC 4437 group in comparison with simulated galaxy groups in the IllustrisTNG50 cosmological simulation.
In \S\ref{sec_discussion}, we discuss environmental properties and the correlations between the magnitude gap, number of member galaxies, and assembly history of low-mass galaxy groups.
%properties of the NGC\,4437 group satellite system in comparison with those of previously studied nearby galaxy groups and simulated galaxy groups in comological simulations (IllustrisTNG50). 
%In particular, we focus on the observed anticorrelation between the magnitude gap and the satellite number for low-mass groups.
In the final section, we summarize our main results.

% ===================================
%     2. Data and Galaxy Sample
% ===================================

\section{Data and Dwarf Satellite Galaxy Survey}\label{sec_data}

We use the $g$- and $i$-band imaging data from the second public data release \citep{aih19} for the Wide layer of the HSC-SSP \citep{aih18}, to search for dwarf satellite candidates around NGC\,4437. The HSC is a wide-field optical imager mounted on the 8.2m Subaru telescope. The survey consists of three layers, Wide, Deep, and UltraDeep with a %{\color{red} 
$5\sigma$ %} 
photometric depth of $r \sim$ 26, $r \sim$ 27, and $r \sim$ 28 mag, respectively. 

The images from the HSC-SSP are fully reduced, sky subtracted, and coadded with the pipeline \texttt{hscPipe} designed for the LSST \citep{bos18,ive19}. In \texttt{hscpipe v6}, astrometric and photometric calibrations were performed against the Pan-STARRS1 DR1 catalog \citep{aih19}.
Sky subtraction was carried out by two steps and \citet{aih19} showed that the technique preserves low surface brightness features better than previous versions of \texttt{hscpipe}. 
The pixel scale of HSC images is 0.168{\arcsec} per pixel. The average seeing is 0.58{\arcsec} in {\it i}-band and 0.77{\arcsec} in {\it g}-band. 

%%%%%%%%%%%%%%%%%%%%%%%%%%%%%%%%%
%% Figure 1
%%%%%%%%%%%%%%%%%%%%%%%%%%%%%%%%%

\begin{figure*}[hbt!]
\centering
\includegraphics[scale=0.38]{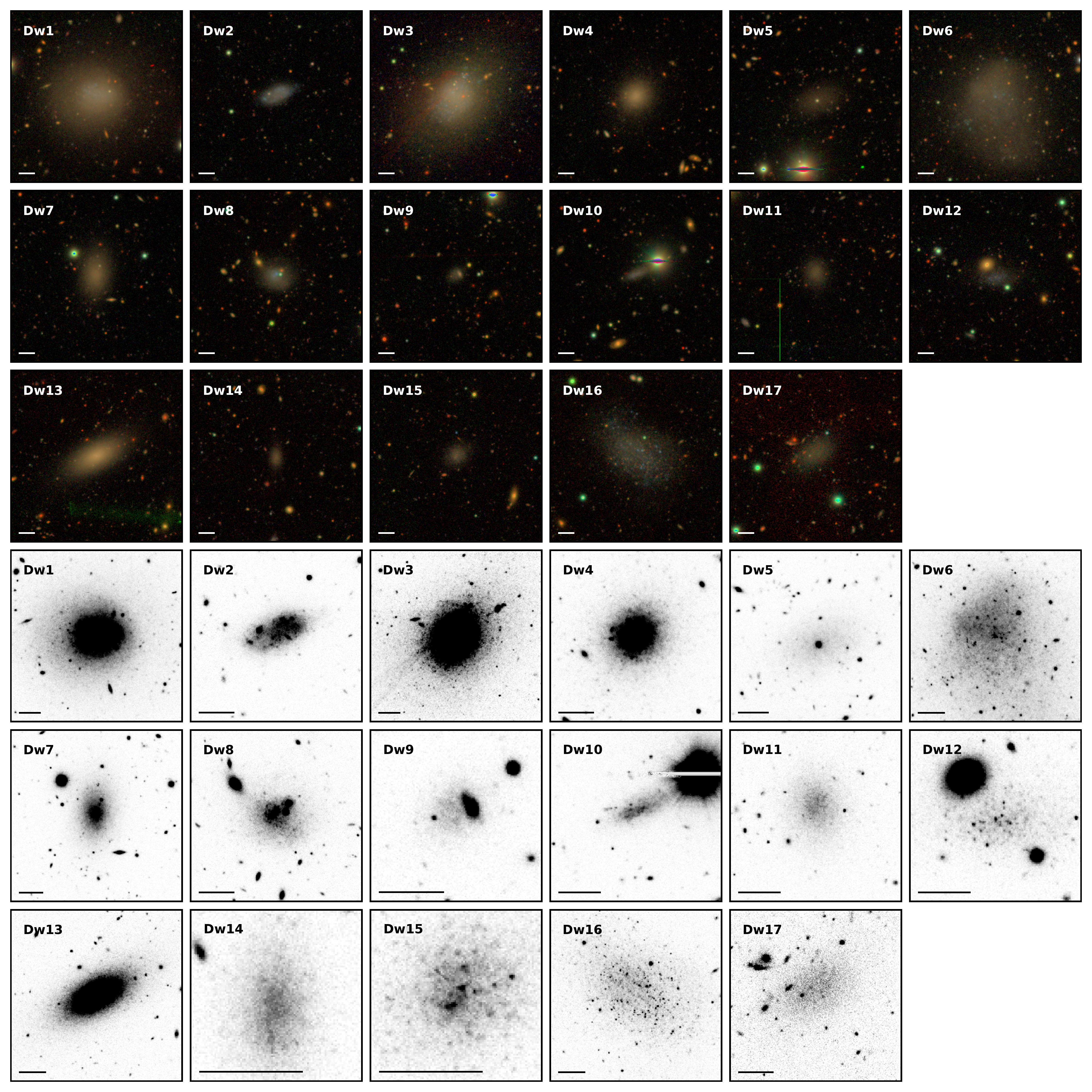} 
%[scale=0.55,{angle=90}]
\caption{(Top) HSC $g$, $r$, $i$ color images of the dwarf satellite candidate galaxies around NGC\,4437. The white bars represent 10{\arcsec} length. North is up, and east to the left. (Bottom) HSC $i$-band zoom-in images of the sample galaxies. The black bars also represent 10{\arcsec} length.}
\label{fig_thumb}
\end{figure*}
%%%%%%%%%%%%%%%%%%%%%%%%%%%%%%%%%

%%%%%%%%%%%%%%%%%%%%%%%%%%%%%%%%%
%% Figure 2
%%%%%%%%%%%%%%%%%%%%%%%%%%%%%%%%%

\begin{figure*}[hbt!]
\centering
\includegraphics[scale=0.7]{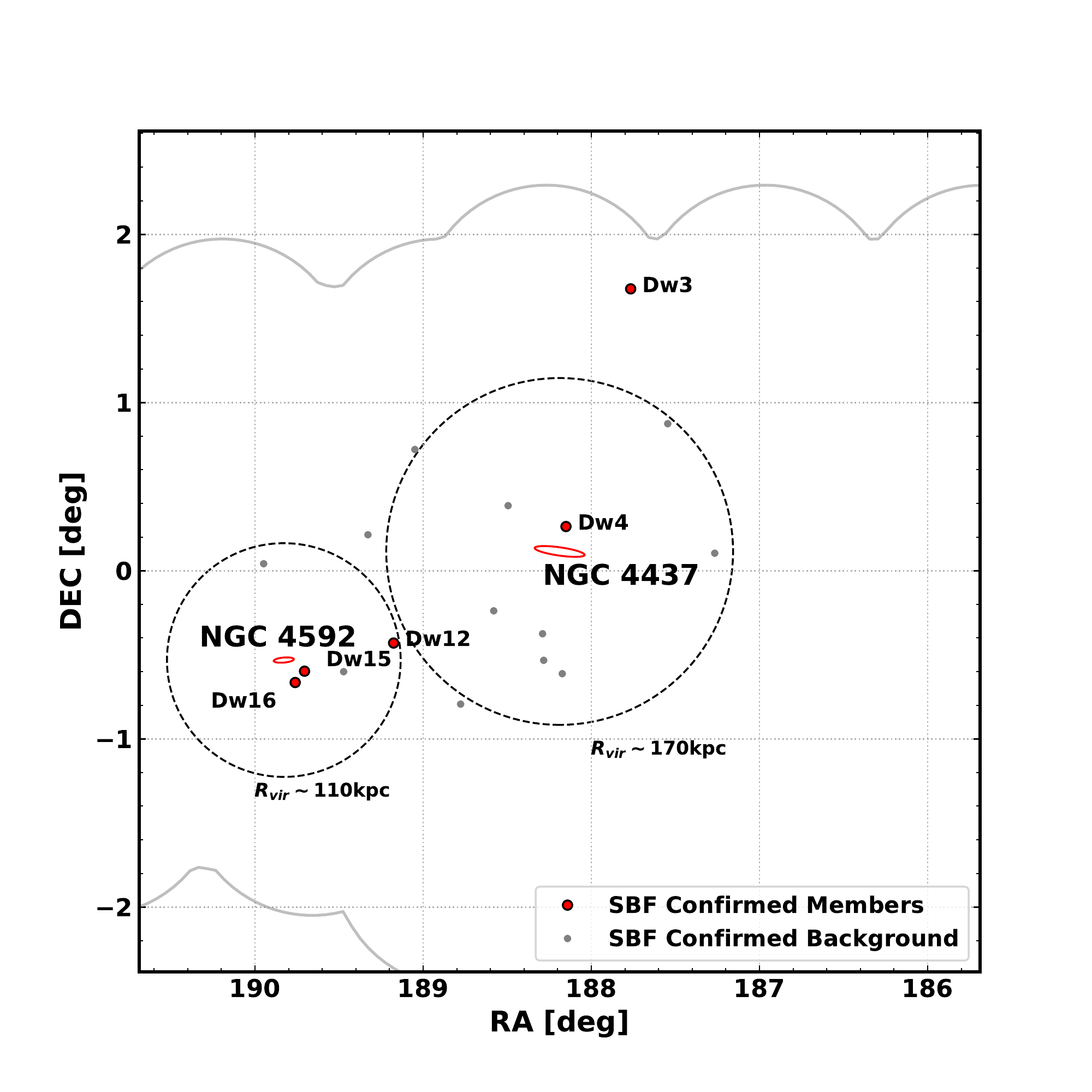} 
%[scale=0.55,{angle=90}]
\caption{Location of dwarf satellite candidate galaxies of the NGC\,4437 group. The two spiral galaxies are shown as red ellipses. The projected virial radii ($R_{vir}$) of the two spiral galaxies are marked as black dashed-line circles.
Red circles denote the group members, and gray dots the background galaxies.
The boundary of HSC $i$-band footprints is displayed as gray lines.} 
\label{fig_FOV}
\end{figure*}
%%%%%%%%%%%%%%%%%%%%%%%%%%%%%%%%%

NGC\,4437 is located 
in %near the center of 
the WIDE12H field, one of the Wide layers of the survey. 
We first visually inspected the 5\textdegree$\times$4\textdegree ($\sim$ 0.8 Mpc $\times$ 0.6 Mpc at the distance of NGC\,4437) area centered on NGC\,4437 to search for extended objects. We restricted our search to galaxies larger than $10\arcsec$ (450 pc at the distance of NGC\,4437) 
%\textbf{
by visual diameter extent %}
because applying the SBF techniques to smaller low surface brightness galaxies would result in an insufficient S/N.
Also, we ruled out galaxies that show visually too small SBFs to be at a distance of about 9 Mpc. 
%\textbf{
%}

Due to the proximity to the Virgo cluster in the sky ($\sim 12.3$\textdegree), some Virgo members may be included in this area. 
Among the objects of which a radial velocity 
can be found in NASA Extragalactic Database (NED),
%is measured, 
we excluded galaxies that have a relative radial velocity with respect to NGC\,4437 $\lvert{\Delta v_{helio}}\rvert > 400$ \kms.
As a result, we selected Dw1, Dw3, Dw4, Dw6, Dw7, Dw12, Dw13, Dw14, Dw15, and Dw16 as our initial sample.
The selected satellite candidates with velocity measurements are Dw1, Dw3, Dw4, Dw6, and Dw13 %Dw1, Dw2, Dw3, Dw4, and Dw7 
(Figure \ref{fig_thumb} and Table \ref{tab_sbf_samples}).
If an object does not have a velocity measurement, we consider 
 it as a satellite candidate. %These are Dw5, Dw6, Dw8, Dw9, and Dw10.

Then we performed a semi-automated detection following \citet{car20a} to detect low surface brightness galaxies %{\bf 
and %study
estimate the completeness of our search. In this process, we added Dw2, Dw5, Dw8, Dw9, Dw10, Dw11, and Dw17 to the sample. %}

As a result, we find 17 candidate galaxies that vary in size, surface brightness, and morphology.  
Their locations are shown in Figure \ref{fig_FOV}.
Table \ref{tab_sbf_samples} lists basic information of our candidate galaxies.
The coordinates and heliocentric velocities are obtained from NASA Extragalactic Database (NED). 
We derived photometric properties of these galaxies from \texttt{AUTO} magnitudes obtained using Source Extractor \citep{ber96} to the HSC images. %\textcolor{red}{
The faintest object in $r-$band is Dw8, %which is about
with $m_r = 19.58\pm0.02$ mag (which would correspond to $M_r \sim -10.3$ mag assuming the distance of the NGC\,4437).
%Note that Dw8 is behind NGC 4437 so that its absolute magnitude will be brighter! %}
%To measure galaxy colors ($g-i$), we used a circular aperture radius of 80 pixels (13\arcsec). 
%\textcolor{red}{
The color range of our candidate galaxies is $0.2 \lesssim (g-i)_0 \lesssim 0.9$. %}

%%%%%%%%%%%%%%%%%%%%%%%%%%%%%%%%%%%%%
%  Figure arttest
%%%%%%%%%%%%%%%%%%%%%%%%%%%%%%%%%%%%%

\begin{figure*}[hbt!]
\centering
\includegraphics[scale=0.9]{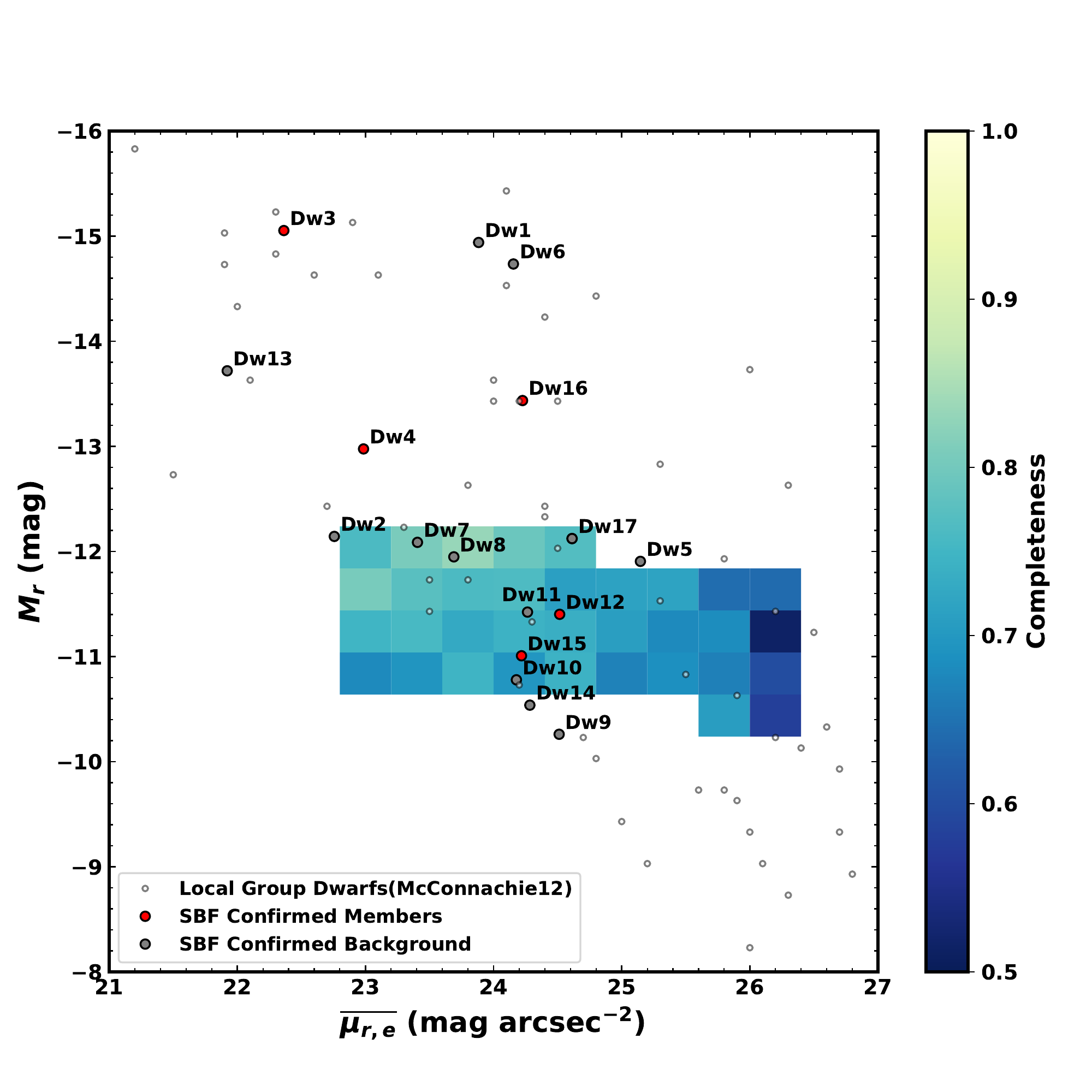} %_arttest.pdf} 
%[scale=0.43, 0.55,{angle=90}]
\caption{Completeness map in $r$-band absolute magnitude vs. average surface brightness within the effective radius, derived from the artificial galaxy test. The color bar represents the derived completeness. %of 
%the sample galaxies are shown as circular symbols, with orange 
Red filled circles and gray filled circles denote confirmed group members and %gray circles being 
background galaxies, respectively. 
%Completeness measured from the artificial galaxy test is shown as colors. 
The Local Group dwarf galaxies \citep{mcc12} are shown as gray open circles for comparison.
%TO ADD legends for filled circles in the lower left corner!
} 
\label{fig_art}
\end{figure*}
%%%%%%%%%%%%%%%%%%%%%%%%%%%%%%%%%

%Figure \ref{fig_art} shows 
%We injected artificial galaxies 
%\textbf{
To check if our survey is complete, we injected 20,000 artificial galaxies of various structural parameters ($-10.5 \lesssim M_r \lesssim -12$ mag and $23 \lesssim \overline{\mu_{r, e}} \lesssim 26$ ${\rm mag}$ ${\rm arcsec^{-2}}$) randomly into the survey field and performed semi-automated detection to find the recovery rate. We assume that our automated detection and visual search are complete for galaxies brighter than this range. % (Figure \ref{fig_art}). 
Figure \ref{fig_art} shows a completeness map in $r$-band absolute magnitude and average surface brightness within the effective radius of our sample galaxies. For comparison, dwarf galaxies in the Local Group from \citet{mcc12} are displayed as gray circles. Locations of our sample galaxies are overlapped with those of the Local Group dwarf galaxies. 
The recovery rates (completeness) are indicated as colors.
In general, completeness decreases as the surface brightness decreases or the total magnitude becomes fainter. 
%For galaxies with faint total magnitudes ($M_r \gtrsim -11$ mag), the completeness decreases as increasing surface brightness ($\overline{\mu_{r,e}} \lesssim 23.5$ ${\rm mag}$ ${\rm arcsec^2}$, because of the decreasing size.
More than 70\% of artificial galaxies with $\overline{\mu_{r, e}} < 25$ mag arcsec$^{-1}$ and $M_r < -11$ mag are recovered.
%In general, completeness decreases as the surface brightness decreases or the total magnitude becomes fainter. 
%For galaxies with faint total magnitudes ($m_i \gtrsim 18.5$ mag), the completeness decreases as increasing surface brightness ($\overline{\mu_{i,e}} \lesssim 23.5$ ${\rm mag}$ ${\rm arcsec^2}$, because of the decreasing size. %Note that the SBF measurement requires enough number of pixels within a galaxy so poor completeness in this regime }%Among our galaxy sample, there is no galaxy in this regime and it would have been inadequate for SBF measurement
%The completeness of our survey is $\sim 70\%$
%}

%%%%%%%%%%%%%%%%%%%%%%%%%%%%%%%%%%%%%
%  Table 1
%%%%%%%%%%%%%%%%%%%%%%%%%%%%%%%%%%%%%

\begin{deluxetable*}{cccccccccc}[hbt!]
\tabletypesize{\footnotesize} %{\scriptsize}
\tablecaption{Basic Information of NGC\,4437 and NGC\,4592 \label{tab_spirals} }
\tablehead{
\colhead{Name} & \colhead{R.A.} & \colhead{Decl.} & \colhead{V$_{helio}$\textsuperscript{a}} & \colhead{D$_{\rm TRGB}$\textsuperscript{b}} &  \colhead{$M_{K_s}$\textsuperscript{c}}&  \colhead{$M_{r}$\textsuperscript{d}}&  \colhead{$M_{*}$\textsuperscript{e}} &  \colhead{$R_{vir}$\textsuperscript{f}} & \colhead{Morphology\textsuperscript{g}} \\
 & \colhead{(deg)} & \colhead{(deg)} & \colhead{(\kms)} & \colhead{(Mpc)} & \colhead{(mag)} & \colhead{(mag)}& \colhead{($M_{\odot}$)}& \colhead{(kpc)} &  
}
\startdata
 NGC\,4437 & $188.189958$ & $+0.115028$ & $1128 \pm 5$ & $9.28 \pm 0.39$ & $-22.54$ & $-20.71$ & $1.27\times10^{10}$ & 167 & SA(s)cd\\ 
 NGC\,4592 & $189.828067$ & $-0.532008$ & $1069 \pm 2$ & $9.07 \pm 0.27$ & $-19.86$ & $-18.23$ & $1.07\times10^{9}$  & 110 & SA(s)dm\\ 
\enddata
\tablecomments{
\textsuperscript{a} From recent measurements in NED. \\
\textsuperscript{b} \citet{kim20} \\
\textsuperscript{c} NGC\,4437: from \citet{jar03}, NGC\,4592: \texttt{AUTO} magnitude we measured from 2MASS images using \texttt{SExtractor}. \\
\textsuperscript{d} NGC\,4437: converted from $M_{V_T^0} =-20.48$ mag \citep[Total face-on magnitude;][]{dev91} using $M_V = M_r + 0.23$ \citep{eng21b}, NGC\,4592: \texttt{AUTO} magnitude measured from HSC images using \texttt{SExtractor}. \\
\textsuperscript{e} Calculated assuming $M_*/L_{K_s} = 0.6M_{\odot}/L_{\odot}$. Note that $M_* ({\rm MW}) = 5 \times 10^{10} M_\odot$ \citep{car21}. \\
\textsuperscript{f} Calculated %ing
using the $R_{200}$ versus $M_{200}$ relation obtained from IllustrisTNG50: $log(M_{200}) = 3 \times log(R_{200}) + 5.03$. \\
\textsuperscript{g} \citet{dev91}.
}
\end{deluxetable*}
%%%%%%%%%%%%%%%%%%%%%%%%%%%%%%%%%%%%%

%%%%%%%%%%%%%%%%%%%%%%%%%%%%%%%%%%%%%
%  Table 2
%%%%%%%%%%%%%%%%%%%%%%%%%%%%%%%%%%%%%

\begin{deluxetable*}{cccccccccccc}[hbt!]
\tabletypesize{\scriptsize}
\tablecaption{A list of NGC\,4437 group dwarf satellite candidates \label{tab_sbf_samples} }
\tablehead{
\colhead{Id} & \colhead{R.A.} & \colhead{Decl.} & \colhead{V$_{helio}$} &  \colhead{$m_{g,0}$}&  \colhead{$m_{r,0}$}&  \colhead{$m_{i,0}$} & \colhead{$(g-i)_0$} & \colhead{$R_e$} & \colhead{$\overline{\mu_{i,e}}$}  & \colhead{t$_{exp}$($g$/$i$)} & \colhead{Morphology} \\
 & \colhead{(deg)} & \colhead{(deg)} & \colhead{(\kms)}  & \colhead{(mag)} & \colhead{(mag)}& \colhead{(mag)}& \colhead{(mag)} & \colhead{(\arcsec)} & \colhead{(mag $\arcsec^{-2}$)} & \colhead{(s)} & 
}
\startdata
Dw1  & $187.266083$  &$0.103806$  & $1198 \pm 13$  &  $15.15 \pm 0.02$ &  $14.90 \pm 0.02$ &  $14.55 \pm 0.02$ &  $0.46 \pm 0.02$  & $27.9$ & $23.5$ & 1170/2430 & Tr\\ 
Dw2  & $187.545510$  &$0.873920$  &   &  $17.84 \pm 0.02$ &  $17.69 \pm 0.02$ &  $17.66 \pm 0.02$ &  $0.23 \pm 0.02$  & $5.6$ & $22.7$ & 1170/1830 & Ir\\ 
Dw3  & $187.765899$  &$1.675697$  & $1105 \pm 5$  &  $15.24 \pm 0.02$ &  $14.78 \pm 0.02$ &  $14.60 \pm 0.02$ &  $0.46 \pm 0.02$  & $14.8$ & $22.2$ & 1290/1060 & Tr\\ 
Dw4  & $188.150546$  &$0.262439$  & 1192  &  $17.41 \pm 0.02$ &  $16.86 \pm 0.02$ &  $16.60 \pm 0.02$ &  $0.81 \pm 0.02$  & $7.0$ & $22.7$ & 1740/2070 & Sph\\ 
Dw5  & $188.172714$  &$-0.612622$  &   &  $18.58 \pm 0.02$ &  $17.93 \pm 0.02$ &  $17.72 \pm 0.02$ &  $0.88 \pm 0.02$  & $14.0$ & $24.9$ & 1590/3060 & Sph\\ 
Dw6  & $188.283101$  &$-0.533046$  & 750  &  $15.58 \pm 0.02$ &  $15.10 \pm 0.02$ &  $14.96 \pm 0.02$ &  $0.52 \pm 0.02$  & $27.8$ & $24.0$ & 1590/2460 & Tr\\ 
Dw7  & $188.289818$  &$-0.375221$  &   &  $18.34 \pm 0.02$ &  $17.75 \pm 0.02$ &  $17.49 \pm 0.02$ &  $0.88 \pm 0.02$  & $6.8$ & $23.1$ & 1740/2260 & Sph\\ 
Dw8  & $188.494603$  &$0.386552$  &   &  $18.33 \pm 0.02$ &  $17.89 \pm 0.02$ &  $17.81 \pm 0.02$ &  $0.54 \pm 0.02$  & $6.3$ & $23.6$ & 1020/1230 & Tr\\ 
Dw9  & $188.580343$  &$-0.238891$  &   &  $20.06 \pm 0.02$ &  $19.58 \pm 0.02$ &  $19.42 \pm 0.02$ &  $0.67 \pm 0.02$  & $4.7$ & $24.4$ & 1170/1830 & Sph\\ 
Dw10  & $188.777153$  &$-0.793541$  &   &  $19.46 \pm 0.02$ &  $19.06 \pm 0.02$ &  $18.94 \pm 0.02$ &  $0.54 \pm 0.02$  & $5.8$ & $24.1$ & 1890/2060 & Ir\\ 
Dw11  & $189.049118$  &$0.720478$  &   &  $18.88 \pm 0.02$ &  $18.41 \pm 0.02$ &  $18.21 \pm 0.02$ &  $0.78 \pm 0.02$  & $6.8$ & $24.1$ & 1590/1460 & Sph\\ 
Dw12  & $189.175590$  &$-0.430357$  &   &  $18.88 \pm 0.02$ &  $18.44 \pm 0.02$ &  $18.28 \pm 0.02$ &  $0.21 \pm 0.02$  & $8.2$ & $24.4$ & 1020/1230 & Ir\\ 
Dw13  & $189.328167$  &$0.213417$  & $863 \pm 11$  &  $16.68 \pm 0.02$ &  $16.12 \pm 0.02$ &  $15.84 \pm 0.02$ &  $0.88 \pm 0.02$  & $8.2$ & $21.6$ & 1890/1060 & Sph\\ 
Dw14  & $189.472708$  &$-0.600737$  &   &  $19.84 \pm 0.02$ &  $19.30 \pm 0.02$ &  $19.22 \pm 0.02$ &  $0.86 \pm 0.02$  & $5.2$ & $24.2$ & 1740/1660 & Sph\\ 
Dw15  & $189.705284$  &$-0.597846$  &   &  $19.18 \pm 0.02$ &  $18.83 \pm 0.02$ &  $18.67 \pm 0.02$ &  $0.62 \pm 0.02$  & $5.2$ & $24.1$ & 2460/1490 & Sph\\ 
Dw16  & $189.760798$  &$-0.665024$  &   &  $16.82 \pm 0.02$ &  $16.40 \pm 0.02$ &  $16.27 \pm 0.02$ &  $0.27 \pm 0.02$  & $17.0$ & $24.1$ & 2460/1090 & Ir\\ 
Dw17  & $189.948333$  &$0.041222$  &   &  $18.28 \pm 0.02$ &  $17.71 \pm 0.02$ &  $17.50 \pm 0.02$ &  $0.79 \pm 0.02$  & $12.0$ & $24.4$ & 1170/1030 & Sph \\
\enddata
\tablecomments{\textbf{Galaxy References.} Dw1: 2dFGRS N320Z113 \citep{met89}, Dw2: APMUKS(BJ) B122737.31+010900.4 \citep{mad90}, Dw3: CGCG 014-054 \citep{zwi61}, Dw4: MGC 34050 \citep{lis03}, Dw5: APMUKS(BJ) B123007.66-002012.9 \citep{mad90}, Dw6: UGCA 285 \citep{kar68}, Dw7: APMUKS(BJ) B123035.83-000559.0 \citep{mad90}, Dw10: SDSS J123506.51-004736.7 (SDSS Data Release 6), Dw11: SDSS J123611.78+004313.7 (SDSS Data Release 6), Dw12: APMUKS(BJ) B123408.34-000920.0 \citep{mad90}, Dw13: MGC 35222 \citep{lis03}, Dw14: SDSS J123753.44-003602.6 (SDSS Data Release 6), Dw15: APMUKS(BJ) B123615.42-001923.4 \citep{mad90}, Dw16: KDG 171 \citep{kar68}, Dw17: MGC 36065 \citep{lis03}}
%\tablecomments{\textbf{Distance References.} (a) \citet{kim20} (TRGB), (b) \citet{kar13a} (Tully-Fish relation)}
\tablecomments{R.A., Decl., and V$_{helio}$ are from NED recent values. Magnitudes are obtained from \texttt{AUTO} magnitude of \texttt{SExtractor} photometry. Magnitude errors include calibration error 0.017 mag \citep{aih19}. $\overline{\mu_{i,e}}$ indicates average $i$-band surface brightness in half-light radius. Morphology indicates dwarf galaxy morphological types classified by our visual inspection, following \citet{kar13b} (Sph for spheroidal galaxies, Ir for irregular galaxies, and Tr for transition types between spheroidals and irregulars.)}

\end{deluxetable*}
%%%%%%%%%%%%%%%%%%%%%%%%%%%%%%%%%%%%%

For the following analysis, we adopt the AB magnitude in the HSC system which is similar to SDSS and CFHT systems \cite[see][]{kim21}. We correct for foreground reddening on each galaxy based on the extinction maps by \citet{sch98} and the conversion coefficients obtained by \citet{sch11}, and use a subscript 0 for corrected values.

% ===================================
%     3. SBF Measurement
% ===================================

\section{Group membership Confirmation Using SBF Distances}\label{sec_sbf}

\subsection{SBF Measurement}

%%%%%%%%%%%%%%%%%%%%%%%%%%%%%%%%%
%% Figure 3
%%%%%%%%%%%%%%%%%%%%%%%%%%%%%%%%%
\begin{figure*}[hbt!]
\centering
\includegraphics[scale=0.71]{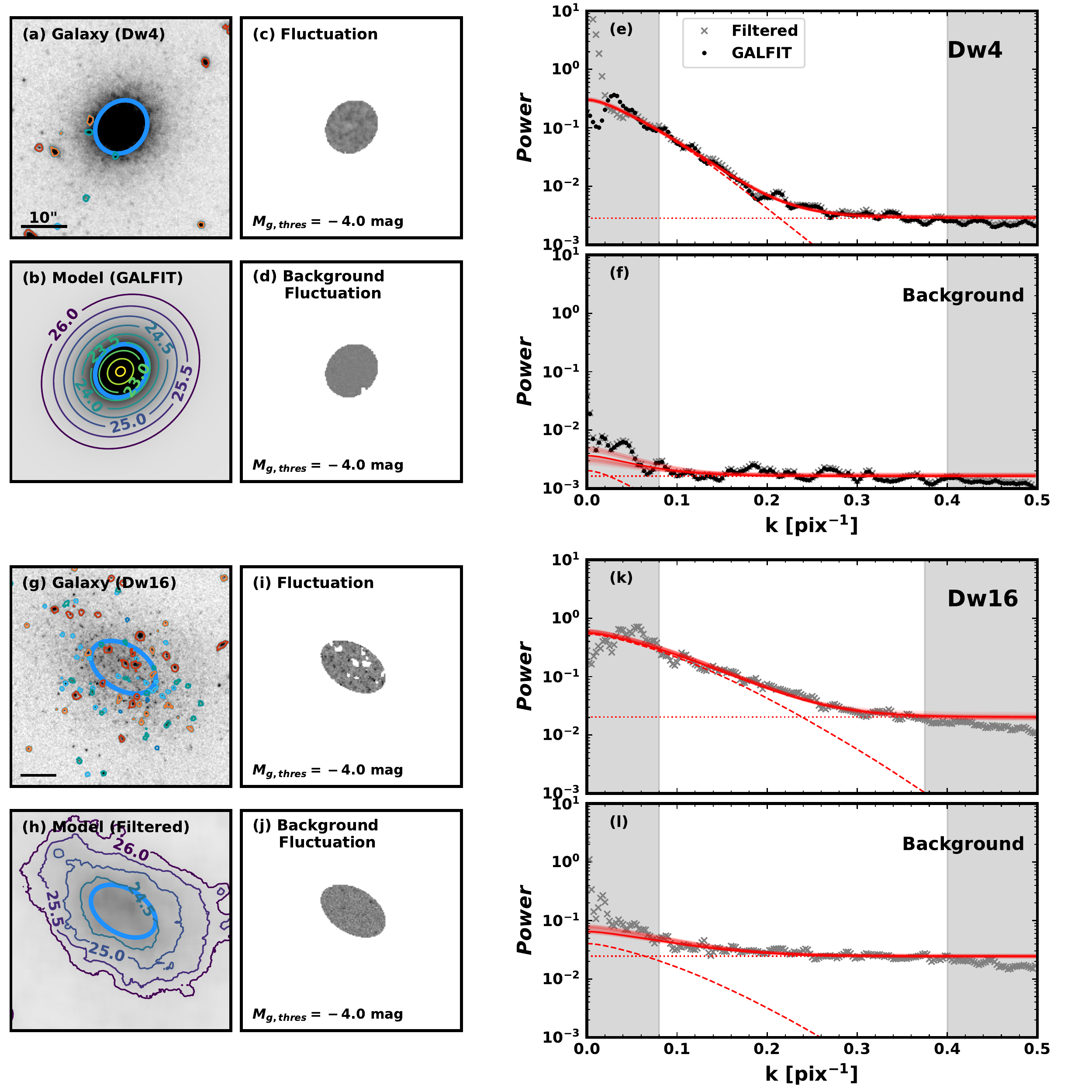} %3.pdf} 
%[scale=0.55,{angle=90}]
\caption{A summary of SBF measurement procedures for Dw4 and Dw16. The upper panels show the case of an early-type galaxy (Dw4) modelled by a single S\'ersic fit, and the lower panels show the case of an asymmetric, late-type, and low surface brightness galaxy (Dw16).
Panels (a) and (g) display $i$-band HSC images with blue ellipses representing the area used for SBF calculation. 
%The dark blue, light blue, cyan, light green, and o5range contours indicate sources masked using thresholds of $M_g = -3.5, -4.0, -4.5, -5.0,$ and $-5.5$ mag, respectively. 
The blue, cyan, teal, orange, and red contours indicate sources masked using thresholds of $M_g = -3.5, -4.0, -4.5, -5.0,$ and $-5.5$ mag, respectively. 
Black bars represent 10{\arcsec} length. Panel (b) shows a smooth galaxy model generated using \texttt{GALFIT} and panel (h) shows a median filtered model image. Contours show the surface brightness as an unit of mag$\arcsec^{-2}$. Panels (c) and (i) are fluctuation images, which are calculated as ${({\rm Galaxy} - {\rm Model})} / {\sqrt{\rm Model}}$. Panels (d) and (j) are examples of background fluctuations sampled from a random area with no apparent bright objects and calculated as $({\rm Background}) / {\sqrt{\rm Model}}$. Panels (e) and (k) show azimuthally averaged power spectra of galaxy fluctuation images and fitted lines. Dashed lines and dotted lines indicate PSF-scale components and constant white noise components, respectively. Panels (f) and (l) show similar power spectra but of background fluctuations.
}
\label{fig_SBF}
\end{figure*}
%%%%%%%%%%%%%%%%%%%%%%%%%%%%%%%%%

While SBF techniques have been widely used for measuring distances to bright early-type semi-resolved galaxies \citep[e.g.,][]{ton88, jer00, mei05, mie07, can18, bla21}, recent studies have shown that SBF techniques are also useful for measuring distances to dwarf galaxies with various morphological types \citep{car19,car21, kim21}. Subtracting the contributions from contaminating sources is trickier in the dwarf regime, but with a statistical approach for background subtraction \citep{car19} and using a $g-$band masking threshold \citep{kim21}, one can derive reliable distances to dwarf galaxies.

In order to estimate the distances to the satellite candidate galaxies and confirm their group membership, we measure SBF magnitudes of the candidate galaxies following the methods described in \citet{kim21}. Here we briefly introduce the main steps of the SBF method and show examples of a dwarf spheroidal galaxy Dw4 %Dw3 
and a blue, asymmetric irregular galaxy Dw16 %Dw10
in Figure \ref{fig_SBF}. 

In a nutshell, the SBF technique measures a PSF-scale fluctuation (dashed-line components in right panels of Figure \ref{fig_SBF}) in Fourier domain of the fluctuation image, ${({\rm Galaxy} - {\rm Model})} / {\sqrt{\rm Model}}$ (Figure \ref{fig_SBF}(c) and (i)), where Galaxy denotes the observed galaxy image (Figure \ref{fig_SBF}(a) and (g)) and Model corresponds to the smooth galaxy light component (Figure \ref{fig_SBF}(b) and (h)). 
For making smooth galaxy models, 
we use \texttt{GALFIT} \citep{pen02} for the galaxies which are well described by a single S\'ersic model or use median filtered images for the galaxies with an asymmetric morphology or star forming regions, with the filter size set to ten times the seeing size, following \citet{kim21}.
%we use \texttt{GALFIT} \citep{pen02} for five galaxies (Dw3, Dw5, Dw7, Dw8, Dw9) which are well described by a single S\'ersic model. 
%For the other five galaxies with an asymmetric morphology or star-forming regions, we use median filtered images with the filter size set to ten times the seeing size, %as
%following \citet{kim21}. 
Using such a filter size, the power spectra calculated using \texttt{GALFIT} and median filtered image (e.g. black dots and gray crosses in Figure (e)) agree very well except for the largest scale ($k<0.05$ pix$^{-1}$), which is not used for the power spectrum fitting.   

We choose an optimal area for SBF calculation of a galaxy to minimize measurement errors.
We measured the SBF using various annular masks (blue ellipses in Figure \ref{fig_SBF}(a) and (g)) and selected an area based on the galaxy's surface brightness and size.
The size, color, and mean surface brightness of the selected areas are shown in Table \ref{tab_sbf_distance}.
For early-type galaxies or low surface brightness galaxies, galaxy colors and fluctuation magnitudes do not depend significantly on the choice of area.
In the case of Dw3, the inner and outer regions of the galaxy show different colors, indicating different stellar populations. Therefore, we divided the galaxy into an inner region and an outer region and calculated their SBFs separately.

For subtracting fluctuations from background sources, namely foreground stars and background galaxies, we measure the fluctuations in multiple background fields as described in \citet{car19}, and subtract the median power. Panels (d) and (j) show examples of background fluctuations and panels (f) and (l) show the corresponding power spectra. In addition, in order to mask young stellar populations which result in stochastic effects, we mask all the sources brighter than a $g-$band masking threshold, $M_{g,{\rm thres}}$. \citet{kim21} showed that the rms of the SBF calibration is the smallest in the case of $M_{g,{\rm thres}}=-4.0$ mag, among the five masking  thresholds $M_{g,{\rm thres}}=-3.5, -4.0, -4.5, -5.0,$ and $-5.5$ mag.
%ranging between $-5.5 < M_{g,{\rm thres}} < -3.5$. 

However, without prior information of distance, it is impossible to know which apparent magnitude corresponds to $M_g= -4.0$ mag in absolute magnitude. Therefore, we try using five thresholds, $m_{g,{\rm thres}}=26.3$,  25.8,  25.3,  24.8, and $24.3$  mag, %These
corresponding to $M_{g,{\rm thres}}=-3.5$, --4.0, --4.5, --5.0, and $-5.5$ mag at the distance of NGC\,4437, $D=9.28\pm0.39$ Mpc derived using the TRGB method \citep{kim20}. 
Note that the absolute masking thresholds change if a galaxy is located at a different distance. For example, at a distance of 11.68 Mpc (7.37 Mpc), the thresholds correspond to $-4.0, -4.5, -5.0, -5.5,$ and $-6.0$ mag ($-3.0, -3.5, -4.0, -4.5,$ and $-5.0$ mag).

Lastly, we use fluctuation -- integrated color relations for dwarf galaxies in the HSC system in \citet{kim21}  to obtain SBF absolute magnitudes and derive distances to the satellite candidate galaxies.
\citet{kim21} derived fluctuation -- integrated color relations for dwarf galaxies  using the $gi$ HSC data for 12 nearby dwarf galaxies of various morphological types. They presented the $i$-band SBF calibrations for different masking thresholds (see their Table 3): 
 $\overline{M_i} = (-2.65\pm0.13)+ (1.28\pm0.24) \times (g-i)_0$,  in the case of $M_{g,{\rm thres}} = -4.0$ mag which leads to the smallest rms scatter of 0.16 mag.

%%%%%%%%%%%%%%%%%%%%%%%%%%%%%%%%%
%% Figure 5 4
%%%%%%%%%%%%%%%%%%%%%%%%%%%%%%%%%
\begin{figure*}[hbt!]
\centering
\includegraphics[scale=0.59]{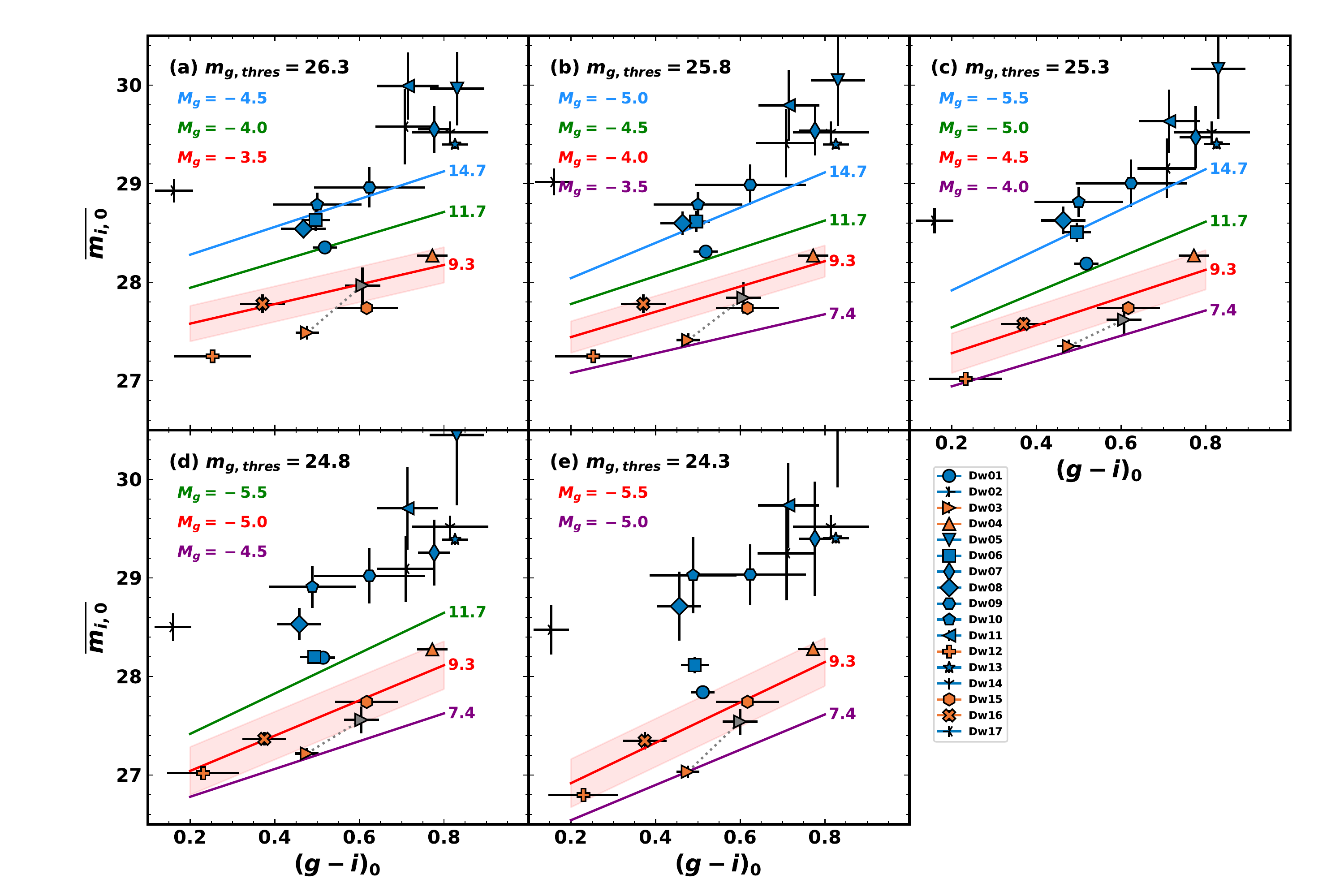} %4.pdf} 
%[scale=0.55,{angle=90}]
\caption{SBF magnitudes of the NGC\,4437 group candidates, with five different $g-$band masking thresholds, $m_{g,{\rm thres}}=26.3, 25.8, 25.3, 24.8,$ and $24.3$ mag. 
Assuming the 9.28 Mpc distance (to NGC\,4437), the thresholds correspond to $M_{g,{\rm thres}}=-3.5,-4.0,-4.5,-5.0,$ and $-5.5$ mag, respectively. Red lines indicate the SBF magnitude -- integrated color relation in \citet{kim21} assuming the 9.28 Mpc distance, for $M_{g,{\rm thres}}=-3.5,-4.0,-4.5,-5.0,$ and $-5.5$ mag, from the left to the right. The red shaded regions represent the rms scatter of the calibrations. 
Five galaxies (Dw3, Dw4, Dw12, Dw15, Dw16) of which SBF magnitudes generally overlap with the shaded red regions are indicated as orange symbols, and the other 12 galaxies as blue symbols. The gray symbol indicates SBF magnitude measured from the outer region of Dw3.
The SBF magnitude -- integrated color relations assuming the 11.68 Mpc distance are shown as orange lines. At the distance of 11.68 Mpc, the masking thresholds correspond to $M_{g,{\rm thres}}=-4.0,-4.5,-5.0,-5.5$ and $-6.0$ mag, respectively. Likewise, the green lines and brown lines correspond to the relations assuming the 14.71 Mpc distance and 7.37 Mpc distance, respectively. 
%{\color{blue} To add Dw IDs somewhere beside symbols. If so, reading the text will be much easier to follow!}
}
\label{fig_threshold}
\end{figure*}
%%%%%%%%%%%%%%%%%%%%%%%%%%%%%%%%%%%%%

%%%%%%%%%%%%%%%%%%%%%%%%%%%%%%%%%%%%%
%  Table 3
%%%%%%%%%%%%%%%%%%%%%%%%%%%%%%%%%%%%%

\begin{deluxetable*}{lccccccc} %[h]
\tabletypesize{\footnotesize}
\tablecaption{SBF Results of NGC\,4437 Group Candidates \label{tab_sbf_distance} }
\tablewidth{0pt}
\tablehead{
\colhead{Name} & \colhead{Annulus} & \colhead{$g - i$} & \colhead{$\overline{\mu}_i$} & 
\colhead{Distance ($D\pm\epsilon_{stat}\pm\epsilon_{calib}$) \textsuperscript{a}}  & \colhead{$M_{g,{\rm thres}}$\textsuperscript{b}} & \colhead{$\sigma_D$\textsuperscript{c}} & \colhead{Membership} \\ & \colhead{(\arcsec)} &\colhead{(mag)} & \colhead{(mag \arcsec$^{-2}$)}& \colhead{(Mpc)} & (mag) & \colhead{(Mpc)} &
}
\startdata
Dw1  & 0.0 - 10.8 &  $0.52\pm0.03$  &  22.6  &  $12.13\pm0.29\pm1.12$ & -4.5 & 0.25 & N\\ 
Dw2  & 0.0 - 11.3 &  $0.16\pm0.04$  &  23.6  &  $23.82\pm1.49\pm2.67$ & -5.0 & 1.48 & N\\ 
Dw3$_{\rm inner}$  & 0.0 - 15.5 &  $0.48\pm0.03$  &  22.4  &  $7.74\pm0.24\pm0.57$ & -4.0 & 0.09 & \multirow{2}{*}{Y}\\ 
Dw3$_{\rm outer}$  & 15.5 - 25.2 &  $0.61\pm0.04$  &  23.7  &  $8.75\pm0.65\pm0.65$ & -4.0 & 0.46 & \\
Dw4  & 0.0 - 6.5 &  $0.77\pm0.04$  &  22.7  &  $9.69\pm0.17\pm0.72$ & -4.0 & 0.05 & Y\\ 
Dw5  & 0.0 - 11.5 &  $0.83\pm0.06$  &  24.8  &  $22.06\pm4.69\pm2.47$ & -5.0 & 0.68 & N\\ 
Dw6  & 0.0 - 14.4 &  $0.50\pm0.03$  &  23.8  &  $15.03\pm0.73\pm1.68$ & -5.0 & 0.31 & N\\ 
Dw7  & 0.0 - 15.8 &  $0.78\pm0.04$  &  24.0  &  $18.21\pm2.12\pm2.04$ & -5.0 & 0.34 & N\\ 
Dw8  & 0.0 - 9.0 &  $0.46\pm0.05$  &  24.0  &  $15.29\pm0.84\pm1.71$ & -5.0 & 0.82 & N\\ 
Dw9  & 0.0 - 5.4 &  $0.62\pm0.13$  &  24.6  &  $16.04\pm1.52\pm1.80$ & -5.0 & 0.42 & N\\ 
Dw10  & 4.3 - 8.6 &  $0.50\pm0.10$  &  24.7  &  $16.20\pm0.96\pm1.81$ & -5.0 & 0.60 & N\\ 
Dw11  & 0.0 - 5.4 &  $0.71\pm0.07$  &  24.2  &  $21.59\pm3.56\pm2.42$ & -5.0 & 1.31 & N\\ 
Dw12  & 0.0 - 7.2 &  $0.25\pm0.09$  &  24.4  &  $8.22\pm0.26\pm0.61$ & -4.0 & 0.39 & Y\\ 
Dw13  & 0.0 - 10.8 &  $0.83\pm0.03$  &  22.4  &  $16.39\pm0.37\pm1.84$ & -5.0 & 0.12 & N\\ 
Dw14  & 0.0 - 5.4 &  $0.81\pm0.09$  &  24.1  &  $17.52\pm0.92\pm1.96$ & -5.0 & 0.12 & N\\ 
Dw15  & 0.0 - 6.1 &  $0.62\pm0.07$  &  24.2  &  $8.31\pm0.16\pm0.61$ & -4.0 & 0.17 & Y\\ 
Dw16  & 0.0 - 10.8 &  $0.37\pm0.05$  &  24.2  &  $9.80\pm0.43\pm0.72$ & -4.0 & 0.37 & Y\\ 
Dw17  & 0.0 - 9.7 &  $0.72\pm0.07$  &  24.2  &  $18.05\pm2.88\pm2.02$ & -5.0 & 1.31 & N\\
\enddata
\tablenotetext{a}{$\epsilon_{stat}$ includes measurement errors, the sum of background subtraction and fitting errors. $\epsilon_{calib}$ includes calibration errors.}
\tablenotetext{b}{Indicates masking threshold %which of 
for the adopted calibration in
\citet{kim21}.} % is used.}
\tablenotetext{c}{Standard deviations of the SBF distances using multiple masking thresholds and calibrations.}
\end{deluxetable*}
%%%%%%%%%%%%%%%%%%%%%%%%%%%%%%%%%%%%%

\subsection{Group Membership Confirmation}

Figure \ref{fig_threshold} shows the SBF magnitudes of the dwarf galaxies, using five different $g-$band masking thresholds. 
We plot our confirmed NGC\,4437 group members as orange symbols and background galaxies as blue symbols. In this subsection, we describe how we decided the membership based on SBF distances.

In each panel, we plot the SBF -- integrated color relations derived by \citet{kim21}, assuming the distance of 9.28 Mpc (to NGC\,4437), as red lines. At the 9.28 Mpc distance, the masking thresholds correspond to $M_{g,{\rm thres}}=-3.5,-4.0,-4.5,-5.0,$ and $-5.5$ mag, respectively. With fainter $M_{g,{\rm thres}}$, the slopes decrease and the y-intercepts increase. The relations assuming smaller and larger distances (7.37 Mpc, 11.68 Mpc, and 14.71 Mpc) and corresponding absolute magnitudes are also shown. If the distance to a galaxy is similar to 9.28 Mpc, its SBF magnitudes are likely to overlap with the shaded red regions. This diagram can be used for roughly estimating distances and choosing which $m_{g,{\rm thres}}$ to use.

The SBF magnitudes do not differ significantly within the error range according to the choice of a masking threshold, for the galaxies in the color range $(g-i)_0\gtrsim0.6$ mag: for example, Dw4, Dw5, Dw7, Dw11, Dw13, Dw14, and Dw17. % are the case. 
The latter six galaxies have significantly fainter SBFs than Dw4. They are even fainter than the relations for $D=14.71$ Mpc. Thus, they are likely Virgo members, not members of the NGC\,4437 group.

In the bluer colors, the SBF magnitudes vary with the choice of a masking threshold. Therefore, deciding the membership of galaxies with bluer colors is more difficult. For instance, it seems that SBF magnitudes of Dw1 and Dw6 are  fainter than the galaxies with orange symbols for $m_{g,{\rm thres}}=26.3$ mag, but the differences decrease at brighter masking thresholds. However, in this case, deciding on a bright masking threshold is not recommended because the thresholds correspond to brighter absolute magnitudes if they are truly located behind the distance of NGC\,4437 and thus young stellar populations might not have been properly masked. We decide that they are likely background objects, based on the fact that SBF magnitudes of Dw1 agree with the relations for 11.68 Mpc and those of Dw6 with 14.71 Mpc in the threshold range $m_{g,{\rm thres}}\geq 25.3$ mag. 

The SBF magnitudes of the Dw3 inner field and Dw12 seem to be slightly brighter than the other three galaxies with orange symbols. If they are located in front of NGC\,4437, it would be better to refer to the cases for brighter masking thresholds than fainter thresholds. The SBF magnitudes of Dw12 overlap with the red shaded regions for the cases of $m_{g,{\rm thres}}=24.8$ and $24.3$ mag, so we decide that it is likely a member of the NGC\,4437 group. For Dw3, while the SBF magnitudes of the inner field shows better agreement with the case of $D=7.37$ Mpc instead of $D=9.28$ Mpc, the outer field shows better agreement with $D=9.28$ Mpc. Thus membership confirmation based on SBF distances is more ambiguous. Fortunately, the velocity of Dw3 ($v_h=1105\pm5$ \kms) is measured to be close to that of NGC\,4437 ($v_h=1128\pm5$ \kms) (see $v_{helio}$ in Table \ref{tab_spirals} and \ref{tab_sbf_samples}) and we conclude that it is a likely member of the NGC\,4437 group.

\begin{figure*}[hbt!]
\centering
\includegraphics[scale=0.6]{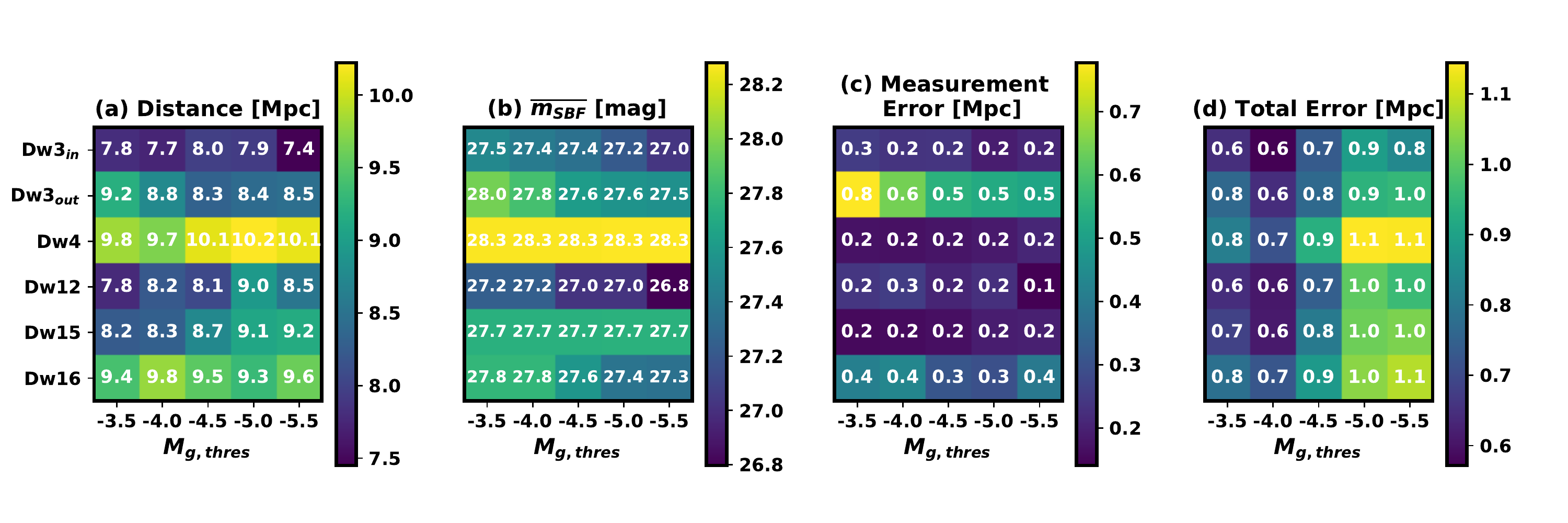} %5.pdf} 
%[scale=0.55,{angle=90}]
\caption{
(a) SBF distances, (b) SBF magnitudes, (c) measurement errors of SBF distances, and (d) total SBF distance errors of the five likely members for five masking thresholds. The values of $M_{g,{\rm thres}}$ are calculated assuming 9.28 Mpc distance. 
}
\label{fig_disterr}
\end{figure*}

Figure \ref{fig_disterr} shows (a) distances, (b) SBF magnitudes, (c) distance measurement errors, and (d) total distance errors of the five likely members for five masking thresholds. 
$M_{g,{\rm thres}}$ is calculated assuming 9.28 Mpc distance. Except for Dw3(outer region) and Dw15, distances do not vary systematically with choice of masking thresholds. In the case of the outer region of Dw3, the systematic trend arises from varying apparent SBF magnitudes.
On the other hand, SBF magnitudes of Dw15 are constant but the difference comes from calibrations. 
In the panel (c), we display the measurement errors in Mpc. Measurement errors are the sum of power spectrum fitting errors and the background subtraction errors. Large errors in 
Dw3(outer region) and Dw16 are due to increased stochasticity in background subtraction in low surface brightness region. In the panel (d), we show total errors, the sum of measurement error and the calibration error. 
Calibration errors dominate total errors. The calibration error is the smallest when using $M_{g,thres} = -4.0$, which is also reflected in total errors. 
We select $M_{g,thres} = -4.0$ for our final choice of distances. 
For Dw3, where measured SBF distance is slightly different when using inner and outer regions, we adopt the average value 8.24 $\pm$ 0.96 Mpc. 

Table \ref{tab_sbf_distance} lists distances calibrated using $m_{g, {\rm thres}} = 25.8$ mag. 
For the five likely members, we assume that the masking threshold corresponds to $M_{g,{\rm thres}}=-4.0$ mag and apply the calibration for such threshold, $\overline{M_i} = (-2.65\pm0.13)+ (1.28\pm0.24) \times (g-i)_0$ (rms = 0.16 mag).
For the other likely background objects, we use calibrations for other thresholds depending on their SBF magnitudes.
$\sigma_D$ indicates standard deviations of SBF distances using multiple masking thresholds. Note that $\sigma_D$ is generally negligible compared to calibration errors.
% and the error ranges are shown in Figure \ref{fig_dist}.

%%%%%%%%%%%%%%%%%%%%%%%%%%%%%%%%%
%% Figure 6  5
%%%%%%%%%%%%%%%%%%%%%%%%%%%%%%%%%
%%%%%%%%%%%%%%%%%%%%%%%%%%%%%%%%%

%%%%%%%%%%%%%%%%%%%%%%%%%%%%%%%%%
%% Figure 6
%%%%%%%%%%%%%%%%%%%%%%%%%%%%%%%%%
\begin{figure*} [hbt!]
\centering
\includegraphics[scale=0.6]{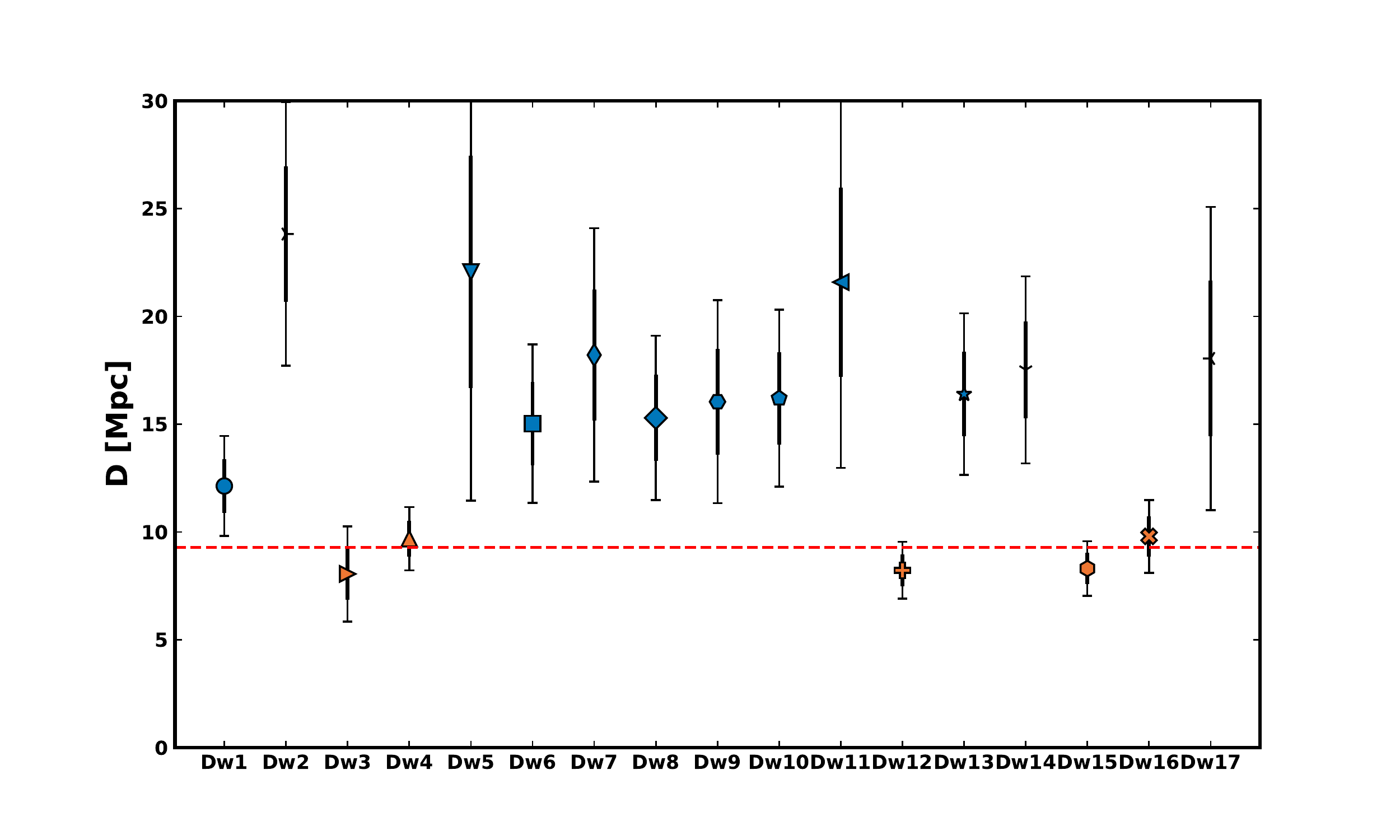} %6.pdf} 
%[scale=0.55,{angle=90}]
\caption{
SBF distances of the satellite candidate galaxies (orange symbols for the group members and blue symbols for the non-group members). Bold and light error bars indicate 1$\sigma$ and 2$\sigma$ uncertainty ranges. The red dashed line denotes the distance of NGC\,4437.
%(b) Spatial distances of the satellite candidate galaxies with respect to NGC\,4437 (filled symbols) and NGC\,4592 (empty symbols). Error bars indicate 1$\sigma$ uncertainty range.
}
\label{fig_dist}
\end{figure*}
%%%%%%%%%%%%%%%%%%%%%%%%%%%%%%%%%

Figure \ref{fig_dist} shows the distances and their error ranges of the dwarf galaxies. While 2$\sigma$ distance range of the five likely members (Dw3, Dw4, Dw12, Dw15, and Dw16) includes the distance of NGC 4437, 9.28 Mpc, that of the other galaxies does not.
% Essentially, the group membership should be decided on the basis of the separation from the host galaxy and the virial radius. 
% The virial radii of NGC\,4437 and NGC\,4592 are approximately 170 kpc (1.0\textdegree) and 110 kpc (0.7\textdegree), respectively. These values are derived by three steps: (1) converting $K_{s}-$band magnitudes to stellar masses assuming $M_*/L_{Ks} = 0.6 M_{\odot}/L_{\odot}$, (2) assuming halo masses using the stellar to halo mass ratio from \citet{beh13}, and (3) using $R_{200}$ versus $M_{200}$ relation obtained from IllustrisTNG50 (see Section \ref{discussion_abundance}) group catalogs. The parameters are listed in Table \ref{tab_spirals}.
% Our distance error ranges are significantly larger than the virial radii, so we only consider the 3-dimensional separation from the spiral galaxies and their error ranges. 
% Figure \ref{fig_dist}(b) shows their 3-dimensional separation from the two spiral galaxies, NGC\,4437 and NGC\,4592. 
% For the five likely members (Dw2, Dw3, Dw6, Dw9, and Dw10), the 1$\sigma$ lower bound of physical separation either from NGC\,4437 or NGC\,4592 includes zero. 
% Thus, we conclude that they are satellites of the NGC\,4437 group.
% On the other hand, for the other five galaxies, the 1$\sigma$ lower bounds do not include zero and their 2$\sigma$ distance lower bounds (panel (a)) exceed the distance of NGC\,4437. 
Adopting the criterion used by \citet{car19a} and \citet{car21}, which is confirming a dwarf as a background object whose $2\sigma$ distance lower bound is beyond the distance of the host, we consider these twelve galaxies as background galaxies.
Note that all the galaxies we newly added by the automated detection are confirmed as background galaxies,  indicating that our initial visual inspection was conservative.
Table \ref{tab_members} lists the confirmed members %candidates 
of the NGC 4437 Group in the order of absolute magnitudes. 

%\textcolor{red}{
Among our likely members, Dw15 has the faintest $r-$band magnitude ($M_r = -11.0$ mag). We assume that our search is complete at least down to $M_r = -11$ mag, among the galaxies with sufficient surface brightness to apply the SBF method.
%}

%%%%%%%%%%%%%%%%%%%%%%%%%%%%%%%%%%%%%
%  Table 4
%%%%%%%%%%%%%%%%%%%%%%%%%%%%%%%%%%%%%

\begin{deluxetable}{lccc} %[h]
\tabletypesize{\footnotesize}
\tablecaption{Confirmed Members %Candidates 
of the NGC\,4437 Group \label{tab_members} }
\tablewidth{0pt}
\tablehead{
\colhead{Name} & \colhead{$M_r$(mag)\textsuperscript{a}} & \colhead{Distance (Mpc)} & \colhead{$V_{helio}$ (\kms)}
}
\startdata
NGC 4437 & $-20.71$ & $9.28\pm0.39$ & $1128\pm5$ \\
NGC 4592 & $-18.23$ & $9.07\pm0.27$ & $1069\pm2$ \\
Dw3 & $-15.05$ & $8.24\pm0.96$ & $1105\pm5$ \\
Dw16 & $-13.44$ & $9.80\pm0.95$ &  \\
Dw4 & $-12.98$ & $9.69\pm0.74$ & $1192$ \\
Dw12 & $-11.40$ & $8.22\pm0.66$ &  \\
Dw15 & $-11.01$ & $8.31\pm0.63$ &  \\
%[-20.7077, -18.23, -15.05, -13.44, -12.98, -11.40, -11.01]
\enddata
\tablenotetext{a}{For the dwarf galaxies, distances are assumed to be at the distance of NGC 4437, 9.28 Mpc.} %Absolute magnitudes of NGC 4437 and NGC 4592 are calculated from apparent magnitudes and TRGB distances. Those of dwarf galaxies are calculated from apparent magnitudes and assuming the distance of NGC 4437, 9.28 Mpc.}
\end{deluxetable}
%%%%%%%%%%%%%%%%%%%%%%%%%%%%%%%%%%%%%

% ===================================
%     4. Satellites Beyond the Virial Radius
% ===================================

\section{Satellites Beyond the Virial Radius}\label{sec_beyond}

In the previous sections, we identified dwarf satellite candidates in about 5\textdegree $\times$ 4\textdegree area around NGC 4437 and found that seven galaxies (NGC\,4437, NGC\,4592, Dw3, Dw4, Dw12, Dw15, and Dw16) are %likely
members of the NGC\,4437 group. 
%%For this reason, 
In this section, we describe the spatial distribution of the galaxies in the NGC 4437 group and discuss it using simulated galaxy groups in IllustrisTNG50.
%%we discuss the properties of the satellites of the NGC\,4437 group in comparison with previous studies on low-mass and MW-mass galaxy groups and with cosmological hydrodynamic simulations.

\subsection{Spatial Distribution of the NGC 4437 Group}

In Figure \ref{fig_FOV}, we plot the spatial distribution of the likely  members of the NGC 4437 Group as red symbols.  
Black dashed circles indicate virial radii of NGC 4437 and NGC 4592, 170 kpc (1\textdegree) and 110 kpc (0.7\textdegree). These values are derived by three steps: (1) converting $K_{s}-$band magnitudes to stellar masses assuming $M_*/L_{Ks} = 0.6 M_{\odot}/L_{\odot}$, (2) assuming halo masses
%\textbf{
$5.3 \times 10^{11} M_\odot$ and $1.5 \times 10^{11} M_\odot$, respectively, %) 
%}
%{\color{red} (TO ADD the value)} 
based on %from using 
the stellar to halo mass ratio from \citet{beh13} 
%\textbf{
(0.024 and 0.007, respectively)
%}
%{\color{red} (TO ADD the value)}
, and (3) %using
deriving virial radii from the $R_{200}$ versus $M_{200}$ relation obtained from IllustrisTNG50 (see Section \ref{discussion_abundance}) group catalogs, $log(M_{200}) = 3 \times log(R_{200}) + 5.03$. The parameters are listed in Table \ref{tab_spirals}. Thus, our field of view covers approximately $5R_{vir} \times 4R_{vir}$ area centered on NGC 4437.
This spatial coverage is wider than most observational studies of nearby galaxy groups, which restrict the group members to those within the projected virial radius \citep[e.g.][]{carlin16, sme18, dav21, car21, mao21}. 

The distance between NGC 4437 and NGC 4592 is about two times the virial radius of NGC 4437.
Dw4 is in the virial volume of NGC 4437 and Dw12, Dw15, and Dw16 are located within the virial radius of NGC 4592. Dw3 is relatively isolated, being outside of the virial radius of both NGC 4437 and NGC 4592. If we restrict the group members to those within the projected virial radius (dashed lines in Figure \ref{fig_FOV}), we should regard NGC 4437 and NGC 4592 as primary galaxies of separate groups. 
%\textbf{However, their observed radial velocities are remarkably similar ($\Delta V_{helio} \sim 60$\kms. }
However, given that their observed radial velocities are remarkably similar ($\Delta V_{helio} \sim 60$\kms), 
%{\bf 
it is likely that they are gravitationally bound. %they are likely gravitationally bound. 
However, the velocity information is not conclusive because we do not know their tangential velocities.
%We do not know the tangential velocities, so we cannot conclude whether they form a single group or separate groups using observed velocities. 
%To address this, 
To decide whether or not to view the two galaxies as consisting a single group, in the next subsection, we check spatial extent and velocity distributions of galaxies in the IllustrisTNG50 mock galaxy groups, especially for if the galaxies located outside the virial radius of the host galaxies are truly gravitationally bound. %and velocities of galaxies in the IllustrisTNG50 mock galaxy groups 

\subsection{IllustrisTNG50 Mock Galaxy Groups}\label{sec_TNG}

% The spatial coverage of most observational studies is restricted to a virial radius from the primary galaxy \citep[e.g.][]{carlin16, dav21, car21, mao21},
% %most observational studies search satellite galaxies within a virial radius from the primary galaxy \citep{car21, mao21}, 
% Thus, we primarily discuss the NGC 4437 group by comparing with simulated galaxy groups for the rest of the paper.
% In this subsection, we describe  the simulated galaxy group sample from IllustrisTNG50.

We use group catalogs from the IllustrisTNG50 project \citep{mar18, nai18, nel18, pil18, spr18, nel19} to select mock galaxy groups. IllustrisTNG is a suite of large-scale cosmological galaxy formation simulations, with three different volumes and resolutions (TNG50, TNG100, and TNG300). 
We choose the smallest resolution TNG50, %considering the resolution. 
which has the baryonic mass resolution 
%The baryonic mass resolution is 
$m_b = 8.5 \times 10^4 M_{\odot}$.
The group catalog includes halos determined by friends-of-friends (FoF) algorithm with a linking length $b=0.2$ and substructures (subhalos) identified by Subfind algorithm \citep{spr01, dol09}.

%Within a FoF group, we define the most massive subhalo as the host galaxy and the other subhalos as satellite galaxies.

\subsubsection{Subhalo Selection} 
\begin{figure*} %[h]
\centering
\includegraphics[scale=0.81]{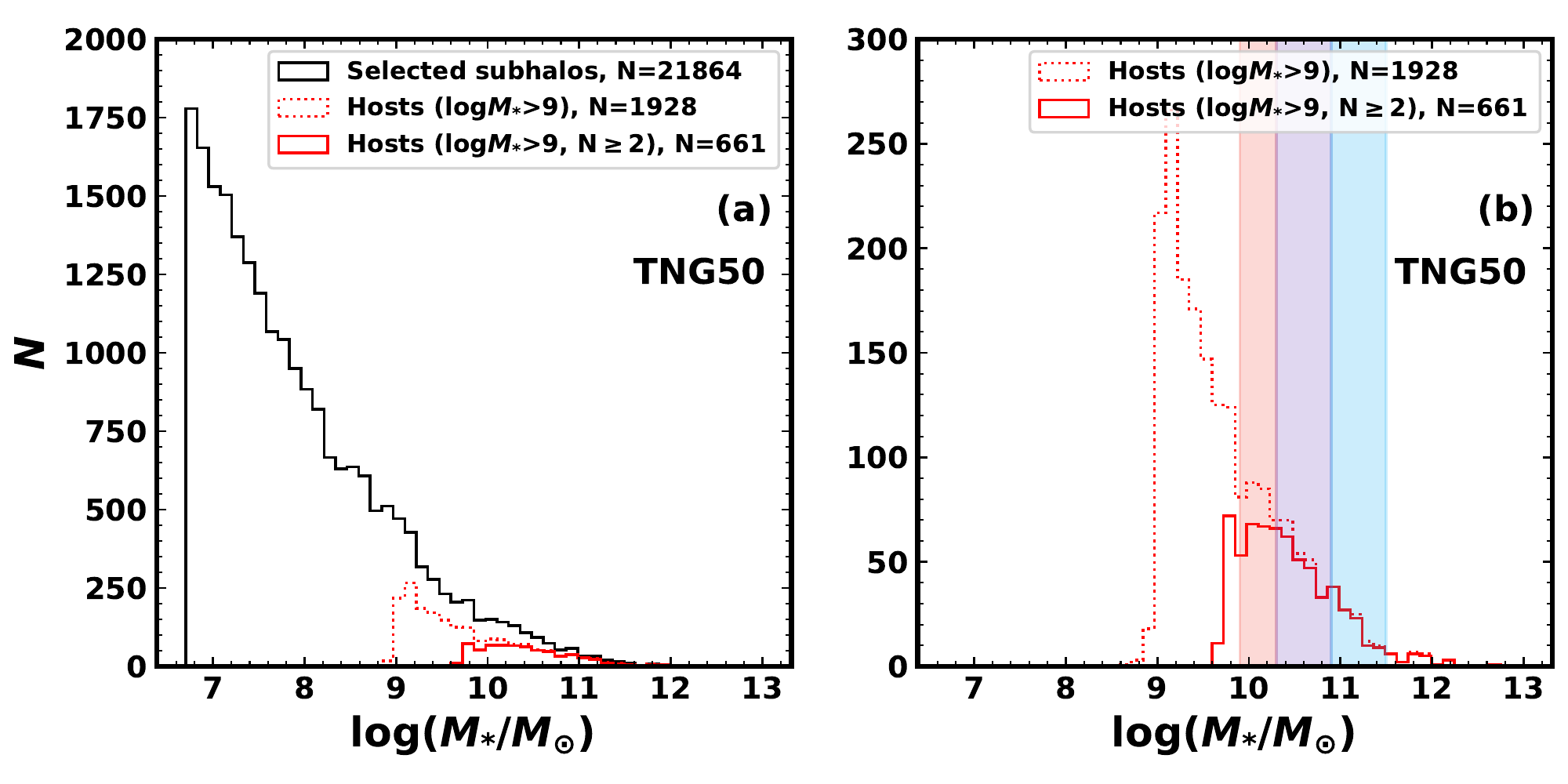} %7.pdf} 
%[scale=0.55,{angle=90}]
\caption{(Left) Stellar mass function of IllustrisTNG50 host subhalos (galaxies) in mock galaxy groups. Black lines display host subhalos with at least $5\times10^{6}M_{\odot}$ within twice the stellar half-mass ratio. Red dotted lines indicate galaxy groups larger than $10^9 M_{\odot}$ and red solid lines show those with at least one more subhalo other than the host subhalo brighter than $M_r=-11$ mag among them. (Right) A zoomed-in version of the left figure. Red, purple, and blue vertical ranges indicate classification of galaxy groups according to host stellar mass: NGC 4437 group-like, MW-like, and massive groups.}
\label{fig_add1}
\end{figure*}

We only consider subhalos with SubhaloFlag = 1 and instantaneous dark matter fraction larger than 10\%, to exclude non-cosmological objects that are fragments or clumps produced through baryonic processes in already formed galaxies \citep{nel19}. 
We also restrict our analysis to luminous subhalos with stellar masses of at least $M_* (<2R_{\rm eff})=5\times10^6{\rm M_{\odot}}$ measured within twice the stellar half-mass radius following \citet{eng21b}, in order to avoid resolution effects. This stellar mass threshold roughly corresponds to $M_r \sim -11$ mag, similar to the completeness limit of our search of NGC\,4437 group satellites. 
In IllustrisTNG50, 21864 subhalos satisfy both criteria.
Black lines in Figure \ref{fig_add1}a shows a stellar mass function of the selected subhalos. 

%%%%%%%%%%%%%%%%%%%%%%%%%%%%%%%%%
%% Figure 7
%%%%%%%%%%%%%%%%%%%%%%%%%%%%%%%%%
%%%%%%%%%%%%%%%%%%%%%%%%%%%%%%%%%

\subsubsection{Galaxy Group Selection}

Next, we select galaxy groups based on two criteria. First, we limit our analysis to FoF groups with a total stellar mass larger than $10^9 M_{\odot}$. In IllustrisTNG50, 1928 FoF groups satisfy this mass limit. Red dotted lines in Figure \ref{fig_add1} shows the host galaxy stellar mass function of such groups.
%A host galaxy is defined as the most massive subhalo among subhalos in a FoF group and ot
Here we define a host galaxy as the most massive subhalo among subhalos in a FoF group and satellite galaxies as the other subhalos. % as satellite galaxies.
Second, we only consider galaxy groups that have at least one satellite galaxy brighter than $M_r = -11$ mag. That is, the galaxy groups with at least one subhalo (other than the host subhalo itself) that satisfies the subhalo selection criteria. Red solid lines in Figure \ref{fig_add1} display the host galaxy stellar mass function of the 661 groups that satisfy both the two criteria. 
Then we classify selected FoF groups according to host stellar mass: low-mass NGC 4437 group-like groups ($9.9 < {\rm log[M_*/M_\odot]} < 10.3$; $N=203$), MW-like groups ($10.3 < {\rm log[M_*/M_\odot]} < 10.9$; $N=237$), and massive groups ($10.9 < {\rm log[M_*/M_\odot]} < 11.5$; $N=96$). 
These are indicated as vertical colored bars in the right panel (Figure \ref{fig_add1}b). 
%We select NGC\,4437-like hosts based on the total mass of stellar particles, ${\rm 9.9 < log[M_{*}/M_{\odot}] < 10.3}$.
%We do not include subhalos residing in a larger FoF group, which are subsystems of a larger structure, in our sample of host galaxies. 
%\textcolor{red}{
%Thus, our selected hosts are field galaxies, not cluster members. %}
%The total number of selected NGC\,4437-like field galaxies is 261. 

%\subsubsection{Spatial Extent of IllustrisTNG50 Mock Galaxy Groups}

\subsubsection{Spatial Extent of Simulated Galaxy Groups}

Now we define two quantities to describe the spatial extent of simulated galaxy groups: $D_{max}$ and $D_{12}$. $D_{max}$ of a galaxy group is the distance between the host galaxy and the farthest satellite galaxy ($M_r < -11$ mag). 
$D_{12}$ of a galaxy group is the distance between the host galaxy and the second brightest member galaxy. 
For the NGC 4437 group, $D_{12} = 0.29^{+0.30}_{-0.03}$ Mpc $\sim 1.7^{+1.8}_{-0.2} R_{200}$ \citep{kim20}. $D_{max}$ is highly uncertain because of the wide distance error ranges in dwarf galaxies.
%For the NGC 4437 group, NGC 4592 has the maximum angular separation with the host galaxy NGC 4437, 1.8\textdegree. This corresponds to a spatial separation of 290 kpc. Therefore, $D_{12} = 290$ kpc $\sim 1.7 R_{200}$ for the NGC 4437 group. 
%For the NGC 4437 group, NGC 4592 has the maximum angular separation with the host galaxy NGC 4437, and the spatial separation is 290 kpc. Thus $D_{12} = 290$ kpc. 

\begin{figure*} %[h]
\centering
\includegraphics[scale=0.81]{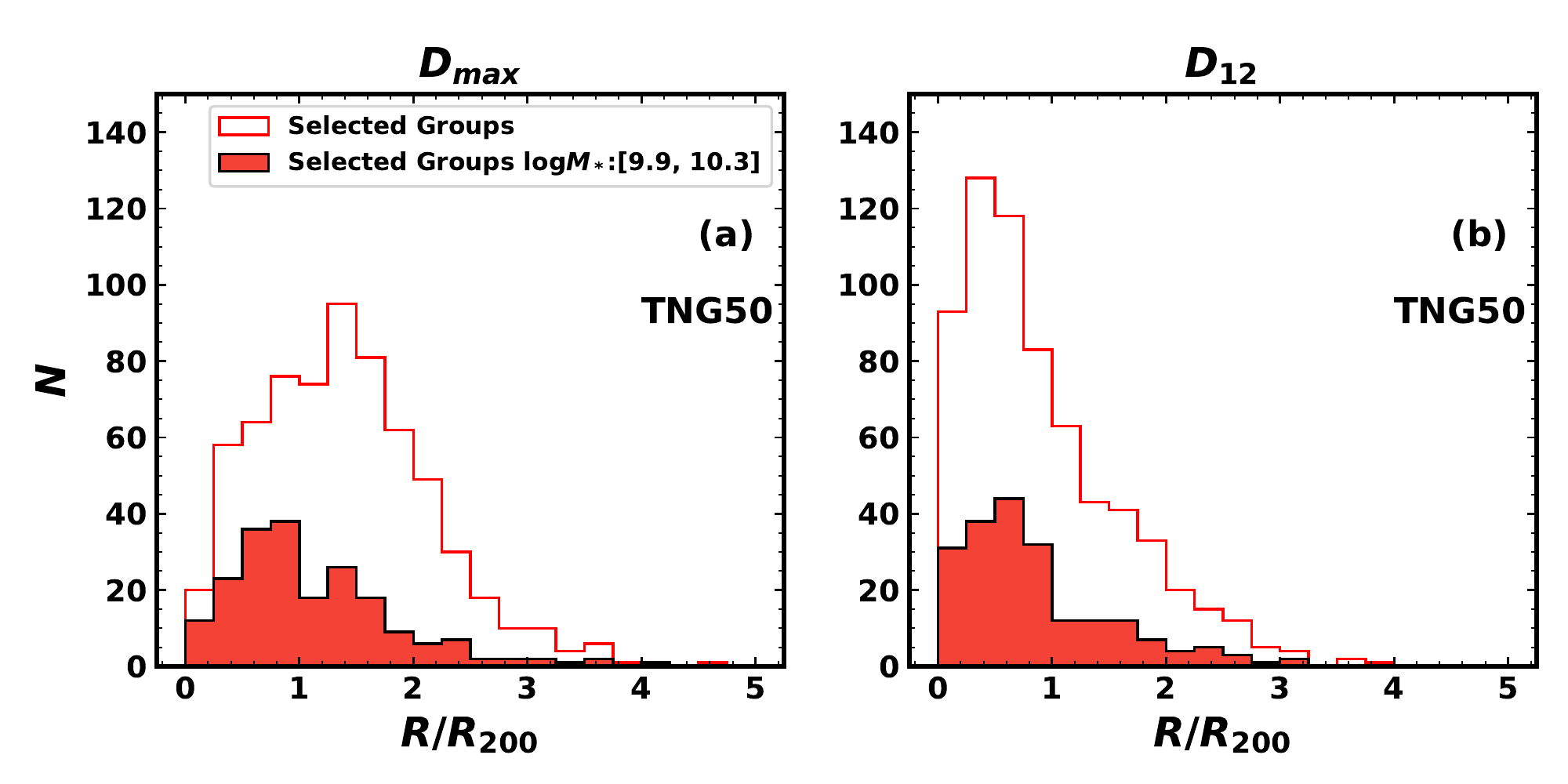} %8.pdf} 
%[scale=0.55,{angle=90}]
\caption{ (Left) $D_{max}$ and (Right) $D_{12}$ distribution of the selected mock galaxy groups in IllustrisTNG50. Red solid lines show all the selected galaxy groups and red filled bars with black edges display NGC 4437 group-like galaxy groups. Note that a significant fraction of satellite galaxies is located outside the virial radius.}
\label{fig_add2}
\end{figure*}

In Figure \ref{fig_add2} we plot the histogram of two quantities in units of the virial radius $R_{200}$. The solid lines indicate all the selected galaxy groups and the filled lines display NGC 4437 group-like galaxy groups, the galaxy groups of which host galaxies have similar stellar masses with NGC 4437 (${\rm 9.9 < log[M_{*}/M_{\odot}] < 10.3}$). 
About half of the NGC 4437 group-like galaxy groups have $D_{max}$ larger than $R_{200}$ and about 10\% of them outside $2R_{200}$. About 30\% of the NGC 4437 group-like galaxy groups have the second brightest member galaxy outside $R_{200}$ and 10\% outside $2R_{200}$. This shows that although a majority of groups have $D_{max}$ and $D_{12}$ within $R_{200}$, a significant fraction of satellite galaxies lie outside the $R_{200}$, like the case for NGC 4437 and NGC 4592. 

To test if these large-distance satellite galaxies are truly bound to the host galaxies, we check their relative velocities.
We find that a majority (75\%) of the second brightest member galaxies of galaxy groups with $D_{12} > R_{200}$ are infalling toward their host galaxies. The other 25\% of them are receding from the host galaxy but have a velocity smaller than the escape velocity. Thus, it is likely that second brightest member galaxies outside the $R_{200}$ will get closer to the primary galaxy and shortly become close satellites. Therefore we conclude that galaxies grouped as a FoF halo in the IllustrisTNG50 are bound, even though their separations are large. 

To summarize, we find that a significant fraction of subhalos in a FoF group in the IllustrisTNG50 simulation are located outside the $R_{200}$ and that they are still bound to the host galaxy. Thus, we decide to view the NGC 4437 group, which has $D_{12} = 1.7 R_{200}$, as a single group. 
%Although we do not know the tangential relative velocity between NGC 4437 and NGC 4592, it is likely that they are bound, given the small radial relative velocity. 
%This gives an important implication for observation: restricting group members to $R_{200}$, one might miss nearby important satellites. %  is easier for observation but one might miss nearby important satellites. %FoF group catalog are 
%Although a majority of galaxy groups have $D_{max}$ and $D_{12}$ smaller than $R_{200}$, 

%\subsubsection{Caveats}
%One caveat in this analysis is that we do not know the tangential velocities of NGC 4437 and NGC 4592. 

%%%%%%%%%%%%%%%%%%%%%%%%%%%%%%%%%
%% Figure 8
%%%%%%%%%%%%%%%%%%%%%%%%%%%%%%%%%
%%%%%%%%%%%%%%%%%%%%%%%%%%%%%%%%%

% ===================================
%     5. Discussion
% ===================================

\section{Discussion}\label{sec_discussion}

%In the previous section, we confirmed seven galaxies (NGC\,4437, NGC\,4592, Dw2, Dw3, Dw6, Dw9, and Dw10) as members of the NGC\,4437 group. 
In this section, we discuss the properties of the satellites of the NGC\,4437 group, %with a particular emphasis on galaxy assembly. %
in comparison with previous studies on low-mass and MW-mass galaxy groups and with cosmological hydrodynamic simulations.

\subsection{Enironmental Quenching of Satellites}

Star formation properties of our satellite galaxies imply that environmental quenching has affected them.
The two satellite galaxies (Dw4 and Dw15) that are located in the shortest projected distance to NGC\,4437 and NGC\,4592,  are  
dwarf spheroidal galaxies with red colors
($(g-i)_0=0.81\pm0.02$ and $0.62\pm0.02$ mag, respectively), implying that they consist of old stellar populations.
On the other hand, the satellites located farther from the pair of spiral galaxies (Dw3, Dw12, and Dw16) are late-types with bluer colors ($(g-i)_0< 0.5$ mag) and ultraviolet flux is detected from GALEX images of them.
%\textbf{
This is consistent with earlier finding that the morphological type of satellite galaxies around the MW, M31, M81 and M101 shows a strong correlation with projected distances from their host galaxies \citep{ein74},  and 
%This 
with color-projected distance relation %has been 
well-known for galaxy clusters \citep{dre80} and the Local Group \citep{van94}.  %}
Recent studies show that similar trends exist even for  low-mass galaxy groups \citep{carlin21, dav21}.
The satellite system of the NGC\,4437 group supports the idea that environmental quenching plays a role even for low-mass galaxy groups where lower ram pressure and tidal fields are expected. 

\subsection{Satellite Luminosity Functions, Number of Member Galaxies, and the Magnitude Gap%and Abundance of the NGC\,4437 Group
}\label{discussion_abundance}

%%%%%%%%%%%%%%%%%%%%%%%%%%%%%%%%%
%% Figure 9
%%%%%%%%%%%%%%%%%%%%%%%%%%%%%%%%%
\begin{figure} %[h]
\centering
\includegraphics[scale=0.65]{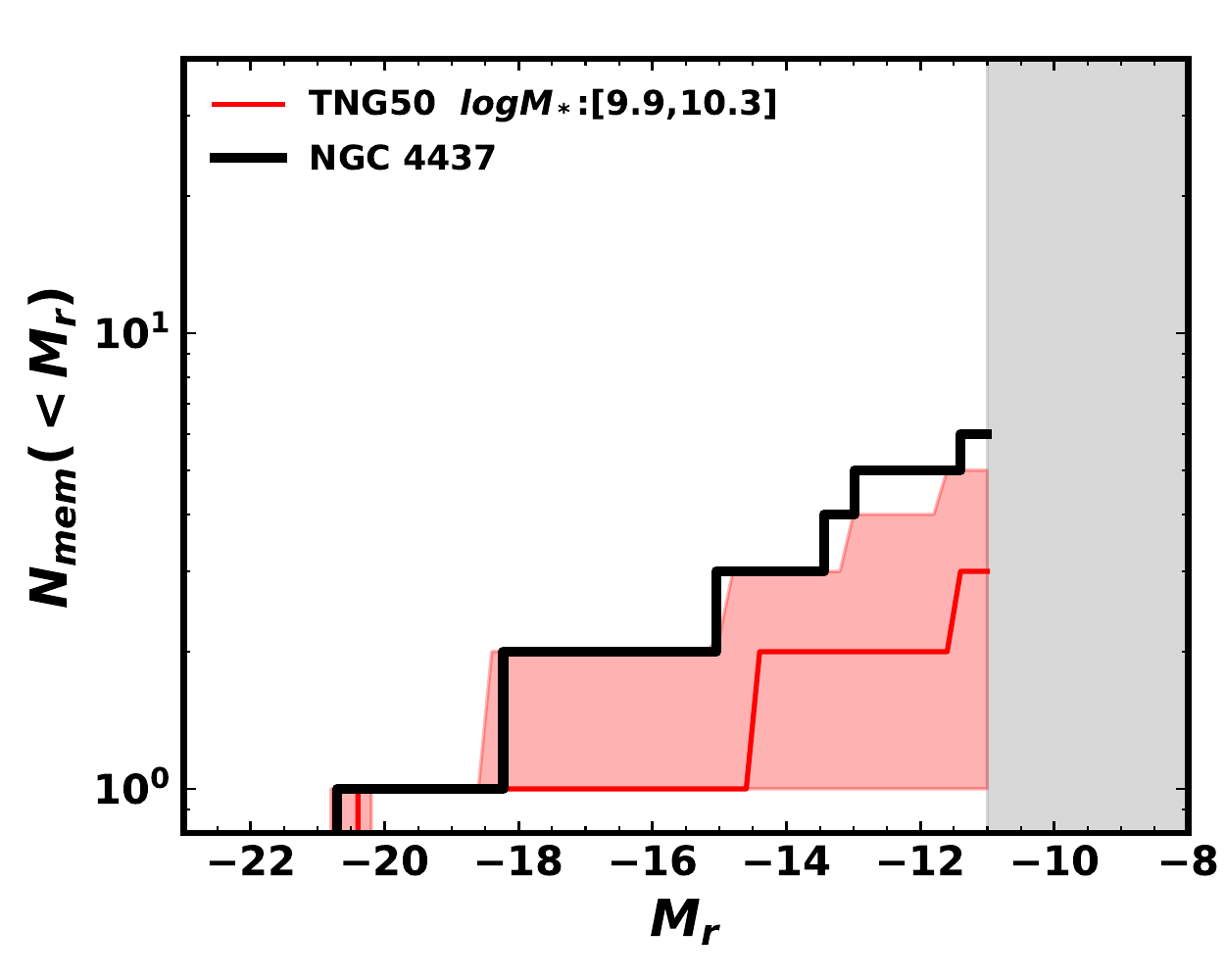} %9.pdf} 
%[scale=0.55,{angle=90}]
\caption{Member galaxy LFs of the NGC\,4437 group (thick black solid line) in comparison with simulated low-mass   groups with $\log[M_{*}/M_{\odot}] = 9.9-10.3$ in IllustrisTNG50.
%the galaxy groups in the LV: 
%MW \citep{mcc12}, M31 \citep{mcc18}, NGC\,4258, NGC\,4631, M51, M101, M94 \citep{car21}, NGC\,628 \citep{dav21}, NGC\,2403 \citep{carlin21}, NGC\,24 \citep{mul20}, and a median LF of 36 MW-like hosts (blue line with $\pm1 \sigma$ band) in the SAGA survey \citep{mao21}. The completeness limit of the satellite search in this study is marked as a gray shaded region ($M_r = -11$ mag). (Right) The same as (a)
%but of the hosts that have a similar stellar mass to that of the NGC\,4437 group: NGC\,4631, M94, and NGC\,618. The red line with $\pm1 \sigma$ band denotes the low-mass groups with $\log[M_{*}/M_{\odot}] = 9.9-10.3$ in IllustrisTNG50.
}
\label{fig_LF}
\end{figure}
%%%%%%%%%%%%%%%%%%%%%%%%%%%%%%%%%

There are seven galaxies ($M_r < -11$ mag) confirmed as members of the NGC\,4437 group.
In this subsection, we compare the spatially complete satellite population of the NGC\,4437 group with the mock galaxy groups in the IllustrisTNG50 cosmological simulation and with other observed galaxy groups.

Figure \ref{fig_LF} displays the observed cumulative luminosity function (LF) of the members in the NGC\,4437 group (thick black solid line) in comparison with the median LF of simulated NGC 4437 group-like galaxy groups (defined in Section \ref{sec_TNG}). %those in mock low-mass groups from the IllustrisTNG50. % and with (b) those in other groups in the literature. 
The median and $\pm1\sigma$ range of the LF of simulated galaxy groups are shown as a red line with a red shaded region.
The photometric limit of this study,
$M_r \sim -11$ mag, is indicated by the gray shaded region.
The NGC 4437 group has a relatively large satellite population compared to the simulated galaxy groups with a similar host stellar mass. 

In addition, the median of the $M_r$ of the second brightest galaxy in the simulated galaxy groups is about $M_r \sim -14.5$ mag. NGC 4592, the second brightest galaxy in the NGC 4437 group, is relatively bright ($-18.2$ mag) among the NGC 4437 group-like simulated galaxy groups. This means that
%Also, 
it has a relatively small \emph{magnitude gap} ($\Delta m_{12} = 2.5$ mag) between the host galaxy (NGC 4437) and the brightest satellite galaxy (NGC 4592) compared to the simulated galaxy groups.

\begin{figure*} %[h]
\centering
\includegraphics[scale=0.75]{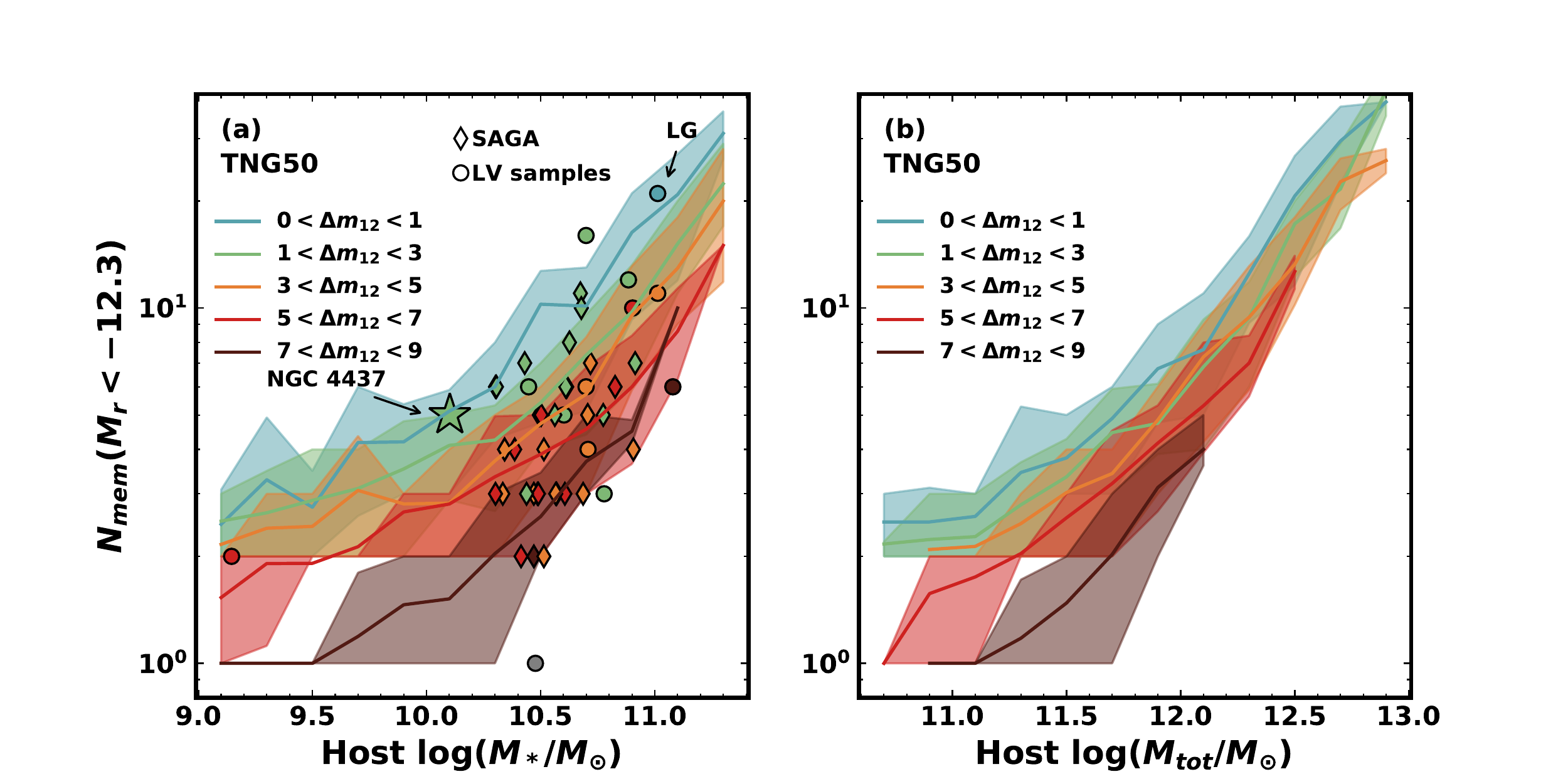} %10.pdf} 
%[scale=0.55,{angle=90}]
 \caption{
 (a) Relations between the number of member galaxies ($N_{mem}$)  brighter than $M_r = -12.3$ mag and stellar mass of the host galaxies from the IllustrisTNG50 simulation. Simulated galaxy groups are classified into five groups by their $\Delta m_{12}$ and their average and 1$\sigma$ range of the $N_{mem}$ are displayed as lines with shaded regions.
 %: the star symbol 
 %the NGC\,4437 group, diamond symbols 
 %the data from the SAGA survey \citep{mao21}, and circle symbols 
 %\citep{car21, mcc12, mcc18, dav21, carlin21, mul20}. 
 %Lines with shaded regions are average and 1$\sigma$ range of the $N_{mem}$  derived from the IllustrisTNG50 simulation, which are divided into five groups by their $\Delta m_{12}$. 
 %The gray circular symbol denotes the $N_{mem}$ of M94, which has $\Delta m_{12} = 9.9$ mag.
 The star symbol indicates the NGC\,4437 group. Diamond symbols and circle symbols show the data from the SAGA survey \citep{mao21} and the LV sample \citep{car21, carlin21, mcc12, mcc18}. Note the stratification of the $N_{mem}$ by $\Delta m_{12}$ for a given host stellar mass. %{\color{red} Where is the star symbol?} 
 (b) Relations between $N_{mem}$ ($M_r < -12.3$ mag) and total mass of the host galaxies derived from the IllustrisTNG50 simulation.
 }
\label{fig_gap}
\end{figure*}
%%%%%%%%%%%%%%%%%%%%%%%%%%%%%%%%%

%%%%%%%%%%%%%%%%%%%%%%%%%%%%%%%%%
%% Figure 11
%%%%%%%%%%%%%%%%%%%%%%%%%%%%%%%%%
\begin{figure} %[h]
\centering
\includegraphics[scale=0.7]{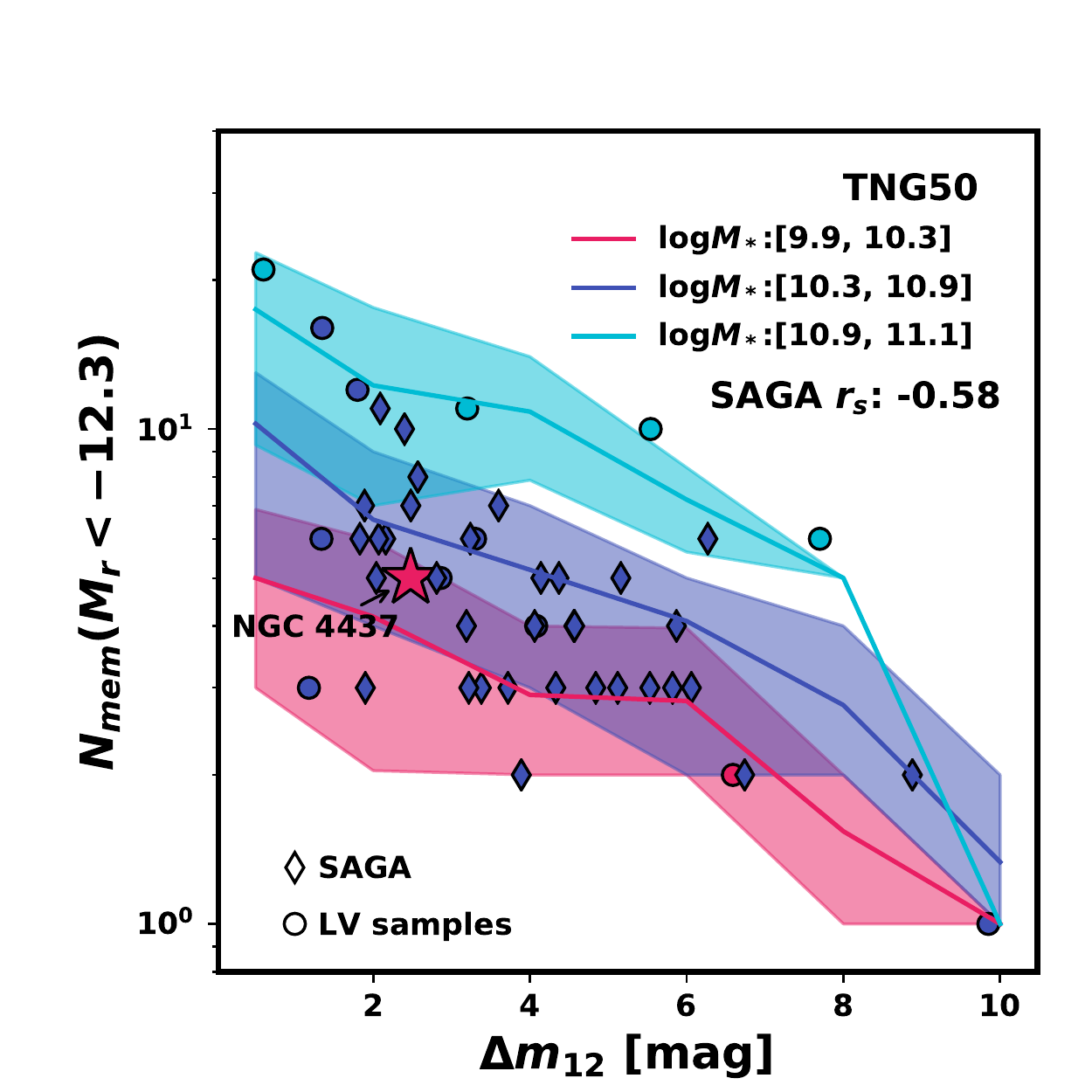} %11.pdf} 
 \caption{
 Relations between $N_{mem}$ ($M_r < -12.3$ mag) and  $\Delta m_{12}$ %total mass of the host galaxies 
 derived from the IllustrisTNG50 simulation, which are divided into
 %five groups by their $\Delta m_{12}$  
 three groups by their stellar mass
 (solid lines with color bands). 
 The star symbol indicates the NGC\,4437 group. Diamond symbols and circle symbols show the data from the SAGA survey \citep{mao21} and the LV sample \citep{car21, carlin21, mcc12, mcc18}.
%{\color{red} Where is the star symbol? TO ADD legends for SAGA and LV in the lower left! }
}
\label{fig_gap2}
\end{figure}
%%%%%%%%%%%%%%%%%%%%%%%%%%%%%%%%%

Previously studied observational evidence of the anticorrelation between the $\Delta m_{12}$ and the number of satellites is based on the comparison between galaxy groups with a very small $\Delta m_{12}$ ($\Delta m_{12} < 0.2$ mag for \citet{hea13} and $\Delta m_{12} < 1$ mag for \citet{wan21}) and the rest.
%The anticorrelation between the $\Delta m_{12}$ and the number of satellites is suggested by
To check whether the anticorrelation extends to a larger gap regime for low-mass galaxy groups,
%To check whether the $\Delta m_{12}$ is related to the number of member galaxies for low-mass galaxy groups, 
we classify the simulated galaxy groups into five groups by $\Delta m_{12}$, $0<\Delta m_{12}<1$, $1<\Delta m_{12}<3$, $3<\Delta m_{12}<5$, $5<\Delta m_{12}<7$, and $7<\Delta m_{12}<9$.

Figure \ref{fig_gap} displays the number of member galaxies of the group, $N_{mem}$ as a function of (a) host galaxy stellar mass and (b) host galaxy total mass from the IllustrisTNG50 simulation. 
Here we consider the satellites brighter than $M_r = -12.3$ mag in order to match the magnitude limit of the SAGA survey \citep{mao21} which will be described below.
The five groups classified by $\Delta m_{12}$ are shown by different colors (the bluer color for the smaller $\Delta m_{12}$).
%The data for the NGC\,4437 group is shown as a star symbol.
%There are 
Two correlations are seen in these figures.
First, the positive correlation between $N_{mem}$ 
and host galaxy stellar mass is seen 
in the simulated groups.
The correlation is tighter against the total mass (including the dark matter halo mass) of the host galaxy.
In addition, the number of satellites from the simulated groups shows a clear stratification in both Figure \ref{fig_gap}(a) and (b): galaxy groups with a smaller $\Delta m_{12}$ have a larger $N_{mem}$. 
Note that the stratification of the same $\Delta m_{12}$ exists across a broad $\Delta m_{12}$ range.

We overplot the observed number of member galaxies and the host stellar mass in Figure \ref{fig_gap}(a) with colored symbols. Colors indicate the five groups of $\Delta m_{12}$. The circle symbols and diamond symbols indicate the LV galaxy groups and the SAGA survey sample, respectively. 
%\textbf{
The LV sample, most of which are from the compilation by \citet{car21}, 
%The LV sample 
consists of 13 galaxy groups of which satellite membership is determined by measuring their distances using either the TRGB or the SBF method:
MW \citep[][for compiled data]{mcc12}, 
M31 \citep[][for compiled data]{mar16, mcc12, mcc18},
the Local Group (LG; including the MW and M31 subgroups and quasi-isolated outlying members classified by \citet{mcc12}),
NGC\,2403 \citep{carlin21},
NGC\,4258 \citep{kim11, spe14, car21},
NGC\,4631 \citep{tan17, car21},
M51 \citep{car21},
M101 \citep{dan17, ben17, ben19, car19a},
M94 \citep{sme18},
NGC\,1023 \citep{tre09, car21},
M104 \citep{jav16, car21},
M81 \citep{chi09, chi13},
and
NGC\,5128 \citep{crn14, cnr19, mul17, mul19}. %}
%Among the 13 groups, nine groups are adopted from the compilation by \citet{car21} (NGC 4258 \citep{kim11, spe14}, NGC 4631 \citep{tan17}, M51, M101, M94, NGC\,1023, M104, M81, and NGC\,5128 (Centaurus A) group). 
%\textcolor{red}{more reference to be added}
The other galaxy groups that are in \citet{car21} compilation but not in this study are excluded 
%The other galaxy groups in \citet{car21} compilation are not included 
either because the spatial coverage is too low (less than 20\% of projected virial area is covered) or the SBF signal-to-noise is too low to confirm satellites as faint as $M_r > -12.3$ mag. 
%The other four groups are NGC 2403 \citep{carlin21}, MW \citep{mcc12}, M31 \citep{mcc18}, and the Local Group (LG) \citep{mcc12, mcc18}. %, which is the sum of the MW and the M31. 
The SAGA survey sample \citep{mao21} consists of galaxy groups with 36 MW-like host galaxies. The MW-like host galaxies are selected by their K-band magnitudes %The K-band magnitudes of host galaxies are selected in 
($-24.6 < M_K < -23$ mag). % range. 
The SAGA survey covered more than 80\% of the projected $R_h=300$ kpc area and confirmed satellite memberships from their 
redshifts. It is considered complete to satellites brighter than $M_r=-12.3$ mag.
%\textcolor{red}{
Note that while the spatial coverage of the LV sample significantly differs from galaxy to galaxy and most of them cover less than the virial area, the SAGA survey sample has a relatively consistent spatial coverage. %}

Most of the references gave luminosities in bandpasses other than $r$-band. We converted the $V$-band magnitudes to $r$-band magnitudes using $M_V = M_r + 0.23$ \citep{eng21b}. For SAGA hosts, we transformed $M_K$ magnitudes to $M_r$ magnitudes using magnitude relations in TNG100 derived using 44 MW-like subhalos with the same $M_K$ range with SAGA selection criteria, $M_r = M_K + 2.48$ (rms = 0.05 mag). The MW-like subhalos are selected using the $g,r,i,z$ magnitudes of the MW \citep{bla16}.

%{\color{red} TO BE REVISED to clarify the meaning! Note that their spatial coverage of the LV sample significantly differs from galaxy to galaxy? and most of them cover less than the virial area.} 
%The SAGA survey sample has a relatively consistent spatial coverage, having more than 80\% of the projected 300 kpc area covered. %and confirmed satellite memberships from their redshifts. It is considered complete to satellites brighter than $M_r=-12.3$ mag.
% The satellite membership is determined by measuring their distances using either the TRGB or the SBF method. Note that their spatial coverage significantly differs. Although  

Although with a large scatter, the observed galaxy groups generally follow the simulated group lines. Also, the positive correlation between the $N_{mem}$ and the host galaxy stellar mass and the anticorrelation between the $N_{mem}$ and the $\Delta m_{12}$ exists for observed galaxy group samples. 

Figure \ref{fig_gap2} illustrates $N_{mem}$ 
as a function of $\Delta m_{12}$ for the same simulated and observed samples as in Figure \ref{fig_gap}. 
To account for host galaxy stellar mass, we sort our compiled sample into three mass groups according to their host stellar masses as defined in Section \ref{sec_TNG}: NGC\,4437-like ($9.9 < {\rm log[M_*/M_\odot]} < 10.3$, pink), MW-like ($10.3 < {\rm log[M_*/M_\odot]} < 10.9$, purple), and massive groups ($10.9 < {\rm log[M_*/M_\odot]} < 11.1$, skyblue). 
A majority of the galaxy groups belong to the MW-like groups.
%We divide and plot the mock groups from the IllustricTNG50 in the similar way (solid lines with shaded regions). %The number of galaxies in each mass group is 306, 241, and 106, respectively.

Among the galaxy groups in the same mass range,
the observational data for the groups show clearly a correlation between the number of satellites and $\Delta m_{12}$:  the smaller the $\Delta m_{12}$, the richer the satellite system. 
The more massive groups have a larger $N_{mem}$ 
for given $\Delta m_{12}$, although the sample numbers in the massive groups and the low-mass groups are small. 
Again, the observed data generally follow the simulated group lines. The Spearman rank correlation coefficient between $\Delta m_{12}$ and $N_{mem}$ 
of the SAGA sample (which is a dominant component of the MW-like groups) is $r_s = -0.58$ ($p-$value = $3.7\times10^{-4}$), which confirms that $\Delta m_{12}$ and $N_{mem}$ are anticorrelated.

We check this correlation in higher resolution simulations, the FIRE project \citep{hop14,hop18}, a suite of high-resolution zoom-in simulation for MW-like galaxies. In their study of the relation between merger history and satellite populations of eight nearby MW-like galaxies, \citet{sme21} defined the stellar mass of the dominant merger ($M_{*,Dom}$, defined as the greater of either total accreted stellar mass ($M_{*,acc}$) or stellar mass of the most massive satellite ($M_{*,Dom.Sat}$)). 
They suggested that $M_{*,Dom}$ shows no clear correlation with the number of satellites ($N_{sat}$ for $M_V < -9$ mag) in the FIRE simulation data,  
while it shows a strong correlation in the observational data (see their Figure 3).
Note that the FIRE data used in their study cover a much smaller mass range of $\log M_{*,Dom}/M_\odot$ ($9-10.3$)  than the observational data ($\log M_{*,Dom}/M_\odot= 8.5-11$).

%%%%%%%%%%%%%%%%%%%%%%%%%%%%%%%%%
%% Figure 12         
%%%%%%%%%%%%%%%%%%%%%%%%%%%%%%%%%
\begin{figure*} [hbt!]
\centering
\includegraphics[scale=0.6]{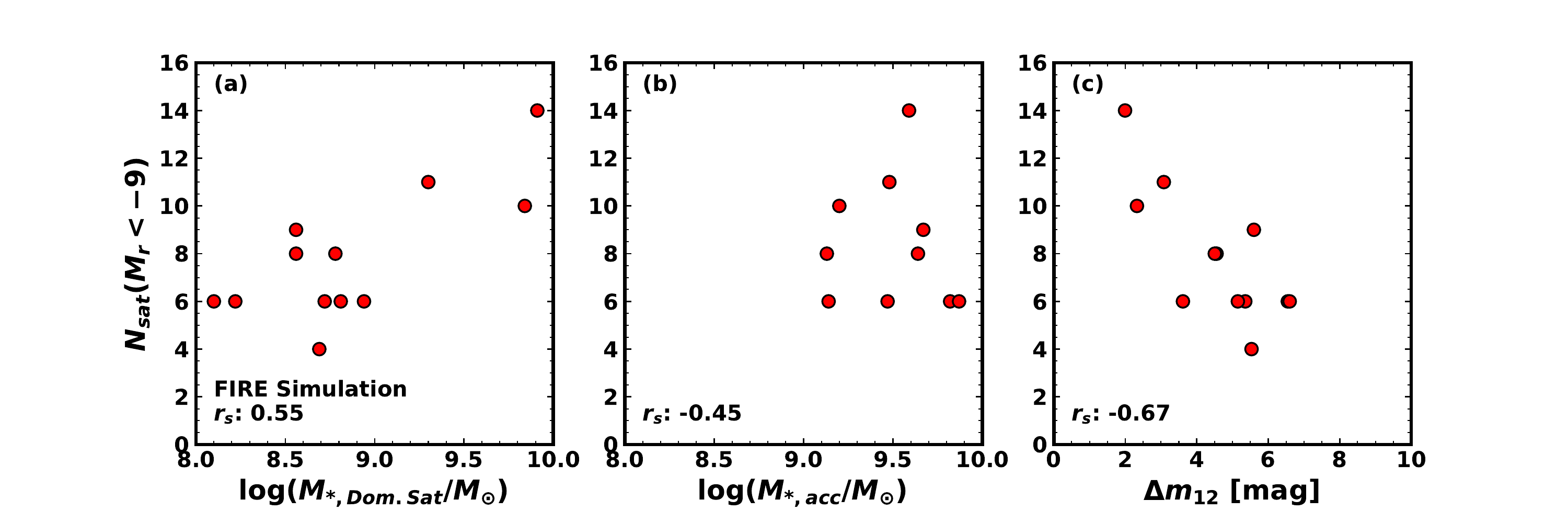} %12.pdf} 
%[scale=0.55,{angle=90}]
 \caption{ Relations of (a) $M_{*,Dom.Sat}$, (b) $M_{*,acc}$, and (c) $\Delta m_{12}$ with $N_{sat} (M_r < -9$), the number of satellites brighter than $M_r = -9$ mag for mock galaxies in the FIRE simulation (data from Table 3 in \citet{sme21}).
Spearman correlation coefficients $r_s$ are given in each panel. Here $\Delta m_{12}$ corresponds to the stellar mass ratio of the host and the most dominant satellite.
 }
\label{fig_FIRE}
\end{figure*}
%%%%%%%%%%%%%%%%%%%%%%%%%%%%%%%%%

Using the information of the FIRE sample in Table 3 of \citet{sme21} we investigate relations between merger mass indicators ($M_{*,Dom.Sat}$ and $M_{*,acc}$) and the number of satellites ($N_{sat}$) in a group. Figure \ref{fig_FIRE}(a) shows a positive correlation ($r_s = 0.55$, $p$-value=0.06) between $M_{*,Dom.Sat}$ and the $N_{sat}$. However, there is a week anti-correlation ($r_s = -0.45$, $p$-value=0.14) between $M_{*,acc}$ and the $N_{sat}$ in Figure \ref{fig_FIRE}(b). As a consequence, $M_{*,Dom}$ shows no correlation with $N_{sat}$ ($r_s=-0.17$, $p$-value=0.59).

If we use the stellar mass ratio between the host galaxy and the most massive satellite galaxy ($M_{*,Dom.Sat}/M_{*,host}$), which is %identical 
equivalent to $\Delta m_{12}$ assuming a constant mass-to-light ratio, we find a %larger
stronger correlation between  $M_{*,Dom.Sat}/M_{*,host}$ and $N_{sat})$ ($r_s=-0.67$, $p$-value=0.02) as seen in Figure \ref{fig_FIRE}(c).
Thus it is shown that the number of satellites increases as the the mass ratio between the host galaxy and the most massive satellite galaxy  decreases (or as their magnitude gap $\Delta m_{12}$ increases).
This result is consistent with those from observational data and IllustrisTNG50 data in this study (as in Figure \ref{fig_gap}).

In summary, both observational data and simulation data show that there is a strong correlation among the three parameters of the galaxy groups: the number of member galaxies is correlated with 
host galaxy stellar mass and with $\Delta m_{12}$.
%magnitude gap.
%, and the number of member galaxies. %Once we know the values of the two among these, we can use them to estimate the value of the remaining parameters. 

\subsection{Magnitude Gap as an Indicator for Galaxy Assembly History}

Since the magnitude gap is correlated with the number of member galaxies 
as well as with host galaxy stellar mass, 
it is likely to also correlate with group assembly histories. If most of the group mass has assembled at an early epoch and no bright galaxy accreted 
since then, the galaxy group might lack bright satellites at present. If a bright satellite existed in the group, it would have been already cannibalized by the host galaxy. Therefore, $\Delta m_{12}$ is likely to be related to mass accretion history. 

This view is supported by the previous studies of fossil groups, which are generally defined as massive systems ($M_{r, host} < -21.5$ mag by \citet{rao14}) with a large magnitude gap ($\Delta m_{12} > 2$ mag). 
The fossil groups are known to have earlier formation times and lack recent satellite accretion \citep{don05, kun17}. However, the correlation between $\Delta m_{12}$ and assembly history for galaxy groups with lower mass than the fossil groups is not well-established.

In this subsection, we investigate host galaxy assembly histories as a function of $\Delta m_{12}$, across a wide host mass range. In particular, we examine cumulative total mass and stellar mass assembly histories, 
halo formation times, and stellar-to-halo mass ratios (SHMRs; the ratio of the total stellar mass to the total mass (including dark matter halos)) of host galaxies in relation to $\Delta m_{12}$. For this purpose we use the group catalogs from IllustrisTNG50, as described in the previous subsection.
From now on, we consider galaxy groups of which the second brightest galaxy is brighter than $M_r = -11$ mag, which is the magnitude limit in our survey of the NGC\,4437 group.
As before, we divide our galaxy groups into three groups according to their host stellar mass: low-mass groups ($9.9 < {\rm log[M_*/M_\odot]} < 10.3$; $N=306$), MW-like groups ($10.3 < {\rm log[M_*/M_\odot]} < 10.9$; $N=241$), and massive groups ($10.9 < {\rm log[M_*/M_\odot]} < 11.5$; $N=101$), and examine their median properties. %}
In each mass group, we divide simulated galaxy groups into five groups according to $\Delta m_{12}$. % similarly.

%%%%%%%%%%%%%%%%%%%%%%%%%%%%%%%%%
%% Figure 13        
%%%%%%%%%%%%%%%%%%%%%%%%%%%%%%%%%
\begin{figure*} %[h]
\centering
\includegraphics[scale=0.85]{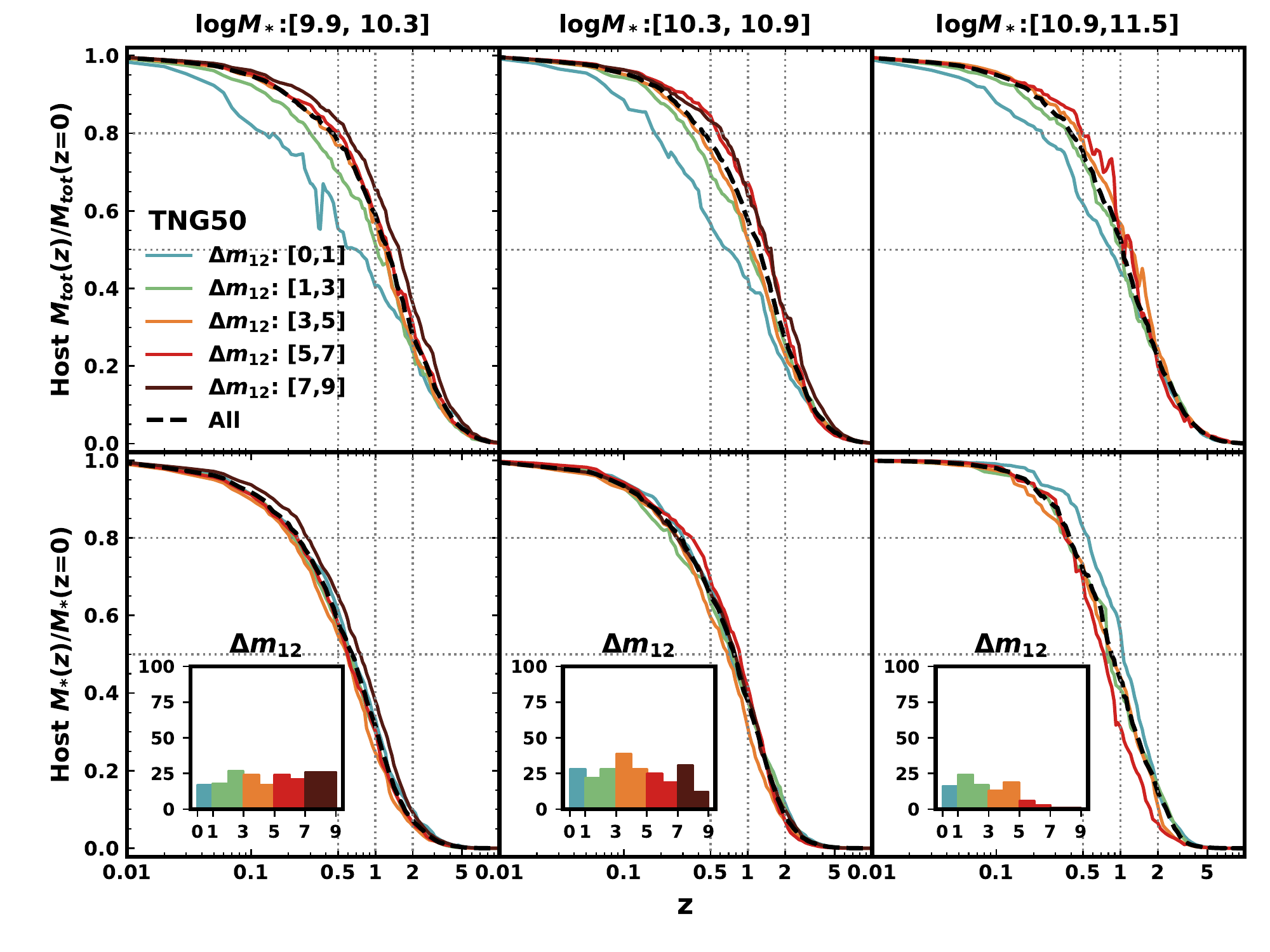} %13.pdf} 
%[scale=0.55,{angle=90}]
\caption{Normalized total mass  (upper panels) and stellar mass (lower panels) %mass
assembly histories of the host galaxies in different stellar mass ranges ($logM_*/M_\odot$: [9.9, 10.3] (low-mass groups), [10.3, 10.9](MW-like groups), and [10.9, 11.5] (massive groups) from left to right, respectively) from the IllustrisTNG50. The median assembly history in each host stellar mass range is indicated as black lines and those divided %by
according to $\Delta m_{12}$ range are shown as colored lines. 
Histograms in the lower panels show the distributions of $\Delta m_{12}$.
In general, galaxy groups with larger $\Delta m_{12}$ tend to have assembled their total mass earlier, while there is no significant trends in stellar mass assembly. 
}
\label{fig_assembly}
\end{figure*}

The upper and lower panels of Figure \ref{fig_assembly} display total mass and stellar mass assembly histories (mass fraction with respect to the current mass as a function of redshift) of the galaxy groups. 
The black lines indicate the median assembly history of all groups.
Less massive groups (left and middle panels) assemble, on average, their total mass at an earlier epoch compared to more massive groups (right panel). 
This tendency is the opposite for the stellar mass assembly history.
Less massive groups assemble their stellar mass at a later epoch, since most host galaxies of low-mass groups show continuous star formation until present day while those of massive groups are quenched.

The distributions of $\Delta m_{12}$ are shown in the histograms of Figure \ref{fig_assembly}. For the MW-like and massive groups, the number of groups decreases with increasing $\Delta m_{12}$. 
%\textcolor{red}{
In contrast, many of the low-mass galaxies have a large $\Delta m_{12}$. %}
%In contrast, the number  of groups in low-mass groups increases  in the large $\Delta m_{12}$ regime. 

We plot the median assembly history of each group divided according to $\Delta m_{12}$ with different colors (the bluer for the smaller $\Delta m_{12}$).
%\textcolor{red}{
Host galaxies %} % The groups
with large $\Delta m_{12}$ assemble their total mass earlier than those with small $\Delta m_{12}$, in all three mass groups.
This suggests that $\Delta m_{12}$ is a useful parameter that distinguishes different total mass assembly histories.
On the other hand, no such trend exists 
for stellar mass assembly histories, as shown in the lower panels.  
%{\color{red}To explain the cause for the difference between total mass and stellar mass? Later or omit!}

%%%%%%%%%%%%%%%%%%%%%%%%%%%%%%%%%
%% Figure 14
%%%%%%%%%%%%%%%%%%%%%%%%%%%%%%%%%
\begin{figure*} %[h]
\centering
\includegraphics[scale=0.58]{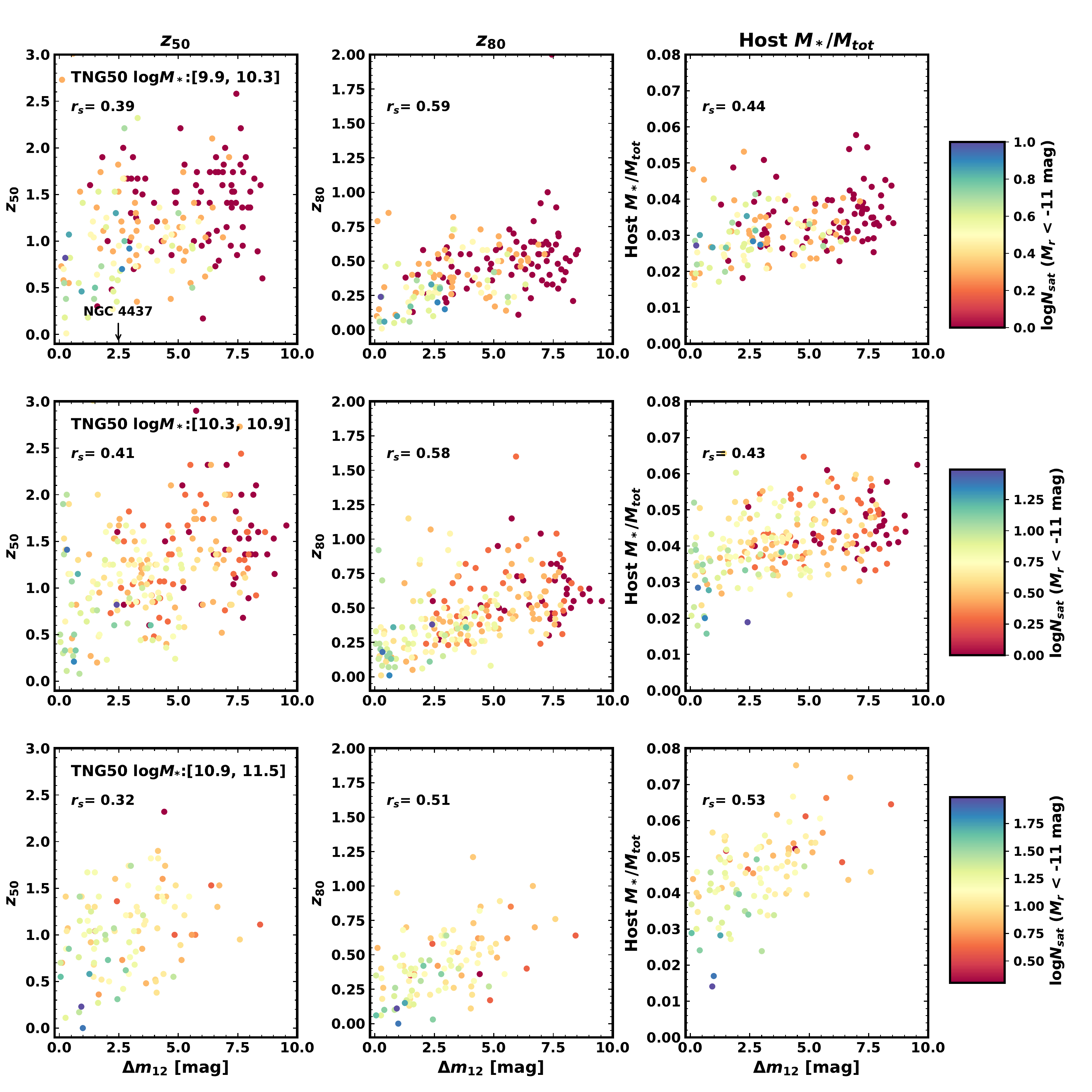} %14.pdf} 
%[scale=0.55,{angle=90}]
\caption{
Halo formation times ($z_{50}$ (left panels) and $z_{80}$ (center panels)) and SHMR (right panels) vs. $\Delta m_{12}$
of the low-mass (top panels), MW-like (middle panels), and massive (bottom panels)  groups
in the IllustrisTNG50. 
Colors denote the number of satellite galaxies brighter than $M_r = -11$ mag. 
The Spearman correlation coefficients ($r_s$) are displayed in each panel. %graph. 
The magnitude gap of NGC\,4437 is annotated as a black arrow on the top left panel.} % and M94 are annotated as black arrows on the top left panel.}
\label{fig_correlation}
\end{figure*}
%%%%%%%%%%%%%%%%%%%%%%%%%%%%%%%%%

Next, we compare halo formation times of groups with different $\Delta m_{12}$. The halo formation time is useful for parameterizing galaxy assembly histories. Here, we examine two parameters, $z_{50}$ and $z_{80}$. $z_{50}$ ($z_{80}$) is a redshift at which the host galaxy assembled 50\% (80\%) of its  mass.

The left panels of Figure \ref{fig_correlation} display relations between $\Delta m_{12}$ and $z_{50}$.
We show three mass groups separately at the top, middle, and bottom panels.
Colors indicate the number of satellite galaxies with $M_r < -11$ mag. There are positive correlations between $\Delta m_{12}$ and $z_{50}$ in all three mass groups, with $r_s \sim 0.4$. Host galaxies of the galaxy groups with a smaller $\Delta m_{12}$ have a larger number of satellites and have assembled their total mass recently. For instance, 80\% of low-mass galaxy groups with $7 < \Delta m_{12} < 9$ have assembled a half of the mass as early as at $z>1$, while only 40\% of low-mass groups with $\Delta m_{12} < 3$ have assembled a half of the mass at $z>1$. 
From this, it is inferred that NGC\,4437 with small magnitude gap ($\Delta m_{12} = 2.5$ mag) is likely to have assembled its mass later than other host galaxies with a large magnitude gap.

The middle panels show the relations between $\Delta m_{12}$ and $z_{80}$.  
The correlations for $z_{80}$ are stronger ($r_s \geq 0.5$) compared to $z_{50}$.
This is consistent with the findings from \citet{kun17} that assembly history of fossil groups and nonfossil groups differs more in recent accretion history and that $z_{80}$ is a more useful parameter than $z_{50}$ in distinguishing them.

The panels in the right represent SHMRs of host galaxies as a function of $\Delta m_{12}$. In all three mass groups, positive correlations exist between the two parameters. The correlation is stronger for more massive groups. 
This is consistent with previous findings that SHMRs of fossil groups are larger than those of nonfossil groups, based on both observations and simulations \citep{har12, kun17}.
For low-mass groups, SHMRs span relatively a narrow range, meaning that halo masses are not much varied given stellar mass. The correlations are weaker and thus $\Delta m_{12}$ does not likely to inform much about SHMRs, not to the same degree as does for higher masses.

To summarize, $\Delta m_{12}$ informs halo formation times and SHMRs of host galaxies even for low-mass groups, in the same manner that can be inferred from the studies of fossil groups: galaxy groups with a large $\Delta m_{12}$ tend to have a small number of satellites, have assembled at an early epoch, and have a large SHMR. 
A large diversity in the number of satellites of nearby galaxies partially originate from diverse assembly histories and an easily observable parameter, $\Delta m_{12}$, can be used as an indicator for assembly times. 
%For instance, the significant difference in the number of satellites between the NGC\,4437 group and the M94 group might be interpreted in terms of galaxy assembly history. The small $\Delta m_{12}$ of the NGC\,4437 group might imply that a half of its mass had assembled as recent as at $z\sim 1$, while the large $\Delta m_{12}$ of the M94 system \citep{sme18} indicates that a half of its mass had assembled as early as at $z\sim 1.5$, indicating that there has been no major accretion of satellites recently. 

%%%%%%%%%%%%%%%%%%%%%%%%%%%%%%%%%
%% Figure 15
%%%%%%%%%%%%%%%%%%%%%%%%%%%%%%%%%
\begin{figure*} %[h]
\centering
\includegraphics[scale=0.7]{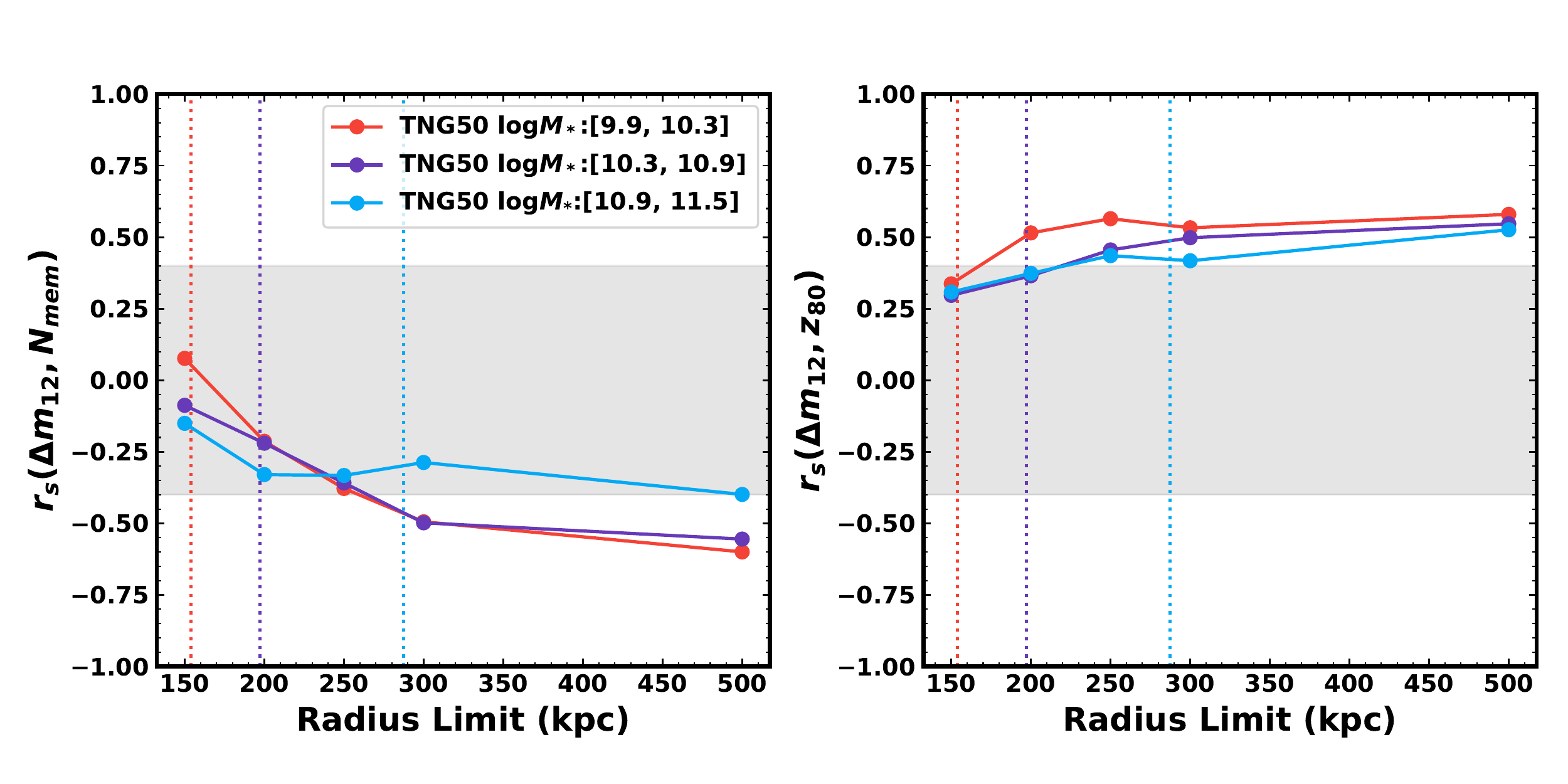} %fig15.pdf} 
%[scale=0.55,{angle=90}]
\caption{Spearman correlation coefficients between $\Delta m_{12}$ and $N_{mem}$ (left) and $\Delta m_{12}$ and $z_{80}$ (right), as a function of radius limits for defining a group. Vertical lines indicate the average virial radius of each groups of mass range. For all the three mass ranges, the correlations increase as radius limit increases.
}
\label{fig_radlimit}
\end{figure*}
%%%%%%%%%%%%%%%%%%%%%%%%%%%%%%%%%

Note that the simulated galaxy groups we use to derive correlations in Figure \ref{fig_correlation} are spatially more complete %\textcolor{red}{
than most observed samples. %}. 
We investigate the relation between the survey coverage and the Spearman correlation coefficients in Figure \ref{fig_radlimit}. We derive Spearman correlation coefficients five times using different radius limits for defining a group, $R_h =$ 150, 200, 250, 300, and 500 kpc. $\Delta m_{12}$ and $N_{mem}$ are determined within each radius limit. %s 
In the left panel we show the correlation coefficients between $\Delta m_{12}$ and $N_{mem}$ ($r_s(\Delta m_{12}, N_{mem})$), and in the right panel $\Delta m_{12}$ and $z_{80}$ ($r_s(\Delta m_{12}, z_{80})$). We plot the correlation coefficients for three different mass ranges with colored circle symbols and the median virial radii as vertical dotted lines. The absolute values of correlations generally increase as radius limit increases, for both panels. 
%An important implication is that the correlations between magnitude gap and the parameters indicating assembly histories may not be significant if the satellite galaxies are searched only in the virial radius. For example, while M94 group is known as an unusually sparse group hosting only two faint dwarfs and the  
The correlations are not significant if the satellite galaxies are searched only in the virial radius. Thus, for $\Delta m_{12}$ to be useful for inferring galaxy assembly histories, it is recommended that satellite galaxies are searched in a wide area as large as $\sim 2R_{vir}$.

\section{Summary}

NGC\,4437 is a low-mass edge-on spiral galaxy ($M_r = -20.7$ mag) 
paired with a dwarf spiral galaxy, NGC\,4592 ($M_r = -18.3$ mag). 
We searched for dwarf galaxies in the 5\textdegree $\times$ 4\textdegree~ area around NGC\,4437 in the WIDE layer of HSC-SSP (Figure \ref{fig_FOV}) and found 17 satellite candidates (Figure \ref{fig_thumb} and Table \ref{tab_sbf_samples}). Our automated detection and visual search are approximately complete down to $M_r = -11$ mag and covers $5R_{vir} \times 4R_{vir}$. 
Then we applied SBF techniques to estimate the distances to the satellite candidates and confirm their membership.
Our main results are summarized as follows.

\begin{enumerate}
    \item \emph{Group membership confirmation.} Based on the SBF distances, we confirm five dwarf  galaxies (Dw3, Dw4, Dw12, Dw15, and Dw16) as members of the NGC\,4437 group. We try measuring the SBF with five masking thresholds for eliminating contaminating sources, $m_{g,{\rm thres}}= 26.3, 25.8, 25.3, 24.8,$ and $24.3$ mag. From the fluctuation --  color diagram, we select five galaxies as likely members and the other twelve %five 
    as likely background galaxies. 
    \item \emph{Environmental quenching for low-mass systems.} The two dwarf galaxies, Dw4 and Dw15, that are located at the shorter projected distance to NGC\,4437 and NGC\,4592, consist of old stellar populations while the other three show star-forming regions. This is consistent with %recent findings \citep{car21, dav21} 
    previous findings that environmental quenching plays an important role for low-mass galaxy groups.
    \item \emph{Satellite richness of the NGC\,4437 group.} Although NGC\,4437 is an order of magnitude fainter than the MW, it has a similar number of satellites %with 
    to those of MW-like galaxy groups. It has a richer satellite system than simulated galaxy groups from IllustrisTNG50, selected by host stellar masses $9.9 < log M_*/M_\odot < 10.3$. However, it has a typical number of satellites when compared with galaxy groups of similar $\Delta m_{12}$, defined as an $r-$band magnitude difference between the first and the second brightest galaxies.
    \item \emph{Magnitude gap and the number of member galaxies.} We find a stratification of the number of member galaxies by $\Delta m_{12}$: the smaller the $\Delta m_{12}$, the larger the number of satellites. This trend is seen both in IllustrisTNG50 (as well as FIRE) simulations and in observed galaxies sample (Figure \ref{fig_gap}).
    \item \emph{Magnitude gap and the galaxy assembly history.} From simulated %\textcolor{red}{
    spatially-complete %} 
    galaxy groups in IllustrisTNG50, we find that galaxy groups (for host galaxy stellar mass range $9.9 < log M_*/M_\odot < 11.5$) with smaller $\Delta m_{12}$ assemble their total mass at a later epoch (Figure \ref{fig_assembly}). Thus, they have relatively later halo formation times $z_{50}$ and $z_{80}$, with $z_{80}$ being more tightly correlated to $\Delta m_{12}$. While SHMR is correlated significantly with $\Delta m_{12}$ for massive groups ($10.9 < log M_*/M_\odot < 11.5$), the correlation gets weaker with decreasing host stellar mass (Figure \ref{fig_correlation}). These findings imply that $\Delta m_{12}$ provides information about assembly histories even for low-mass groups and the diverse assembly histories may account for a large scatter in observed satellite number of nearby galaxy groups. 
    %{\bf 
    In addition, the correlations increase as the radius limit for group definition increases, showing that satellite galaxies should be searched in a wide area to use $\Delta m_{12}$ as an indicator for galaxy assembly. %}
    %The NGC\,4437 group is likely to have assembled its mass later than the M94 group, which has a sparse satellite system with a large $\Delta m_{12}$.
\end{enumerate}

\acknowledgments

This work was supported by the National Research Foundation grant funded by the Korean Government (NRF-2019R1A2C2084019).
J.K. was supported by the Global Ph.D. Fellowship Program (NRF-2016H1A2A1907015) of the National Research Foundation.
We thank the anonymous referee for providing helpful comments. 
We thank Brian S. Cho for his help in improving the English in this manuscript.

%\clearpage

\clearpage

\end{document}